\documentclass[12pt]{article} 
\usepackage{jheppub}

\usepackage{amsmath,amsfonts,amsbsy,amssymb,array,accents,dsfont}
\usepackage{enumerate}
\usepackage[shortlabels]{enumitem}
\usepackage{graphicx} 
\usepackage{caption} 
\usepackage{subcaption} 
\usepackage{amssymb}
\usepackage{upgreek}
\usepackage{bm}
\usepackage{slashed}

\let\includefigures=\iftrue
\let\useblackboard=\iftrue
\newfam\black

\PassOptionsToPackage{ 
     colorlinks=true,
     linkcolor=myBlue,
     urlcolor=blue,
     filecolor=black,
     citecolor=myRed,
}{hyperref}

\usepackage{xparse}
\usepackage{xspace}

\NewDocumentCommand\eqn{om}{%
  \IfNoValueTF{#1}
     {\[ #2 \]}
     {\begin{equation}\label{#1} #2  \end{equation} \expandafter\newcommand\csname #1\endcsname{\eqref{#1}\xspace}\ignorespaces}
}
\NewDocumentCommand\eqna{om}{%
  \IfNoValueTF{#1}
    {\begin{align*} #2 \end{align*}}
    {\begin{equation}\label{#1}\begin{split} #2  \end{split}\end{equation} \expandafter\def\csname #1\endcsname{\eqref{#1}\xspace}\ignorespaces}
}

\newcommand{\rcite}{\cite}

\def\sigmab{{\boldsymbol\sigma}}

\def\sl{\text{sl}}

\def\sltwo{\ensuremath{SL(2,\bR)}}

\def\uone{U(1)}

\def\tight#1{\! #1 \!}  

\def\({\left(}
\def\){\right)}
\def\[{\left[}
\def\]{\right]}

\def\ie{{i.e.}}
\def\eg{{e.g.}}

\def\etc{{etc}}

\def\CFT{{\sst\rm\! CFT}}
\def\BH{{\sst\rm\! BH}}

\def\tot{{\rm tot}}

\def\sfA{{\mathsf A}}		\def\sfB{{\mathsf B}}		\def\sfC{{\mathsf C}}

\def\sfM{{\mathsf M}}						
		\def\sfR{{\mathsf R}}

\def\sfa{{\mathsf a}}		\def\sfb{{\mathsf b}}		\def\sfc{{\mathsf c}}

\def\sfm{{\mathsf m}}						
		\def\sfr{{\mathsf r}}

\DeclareMathSymbol{\medhatsym}{\mathord}{largesymbols}{"62} 

\DeclareMathSymbol{\medtildesym}{\mathord}{largesymbols}{"65}

\def\One{{\hbox{1\kern-1mm l}}}

\def\barray{\begin{array}}
\def\earray{\end{array}}
\def\be{\begin{equation}}
\def\ee{\end{equation}}
\def\bea{\begin{align}}
\def\eea{\end{align}}
\def\nn{\nonumber}

\newcommand{\bR}{{\mathbb R}}
\newcommand{\bS}{{\mathbb S}}
\newcommand{\bT}{{\mathbb T}}

\definecolor{cardinal}{rgb}{0.6,0,0}
\definecolor{darkgreen}{rgb}{0,0.4,0}
\definecolor{green}{rgb}{0,0.4,0}
\definecolor{golden}{rgb}{0.92, 0.7, 0}
\definecolor{midnight}{rgb}{0, 0, 0.5}
\definecolor{darkblue}{rgb}{0, 0, 0.7}

\numberwithin{equation}{section}

\mathchardef\mhyphen="2D

 \def\cB{\mathcal {B}} 
  
 \def\cH{\mathcal {H}} 
  
  \def\cO{\mathcal {O}}

\def\one{{\hbox{\kern+.5mm 1\kern-.8mm l}}}
\def\zero{{\hbox{0\kern-1.5mm 0}}}

\newcommand{\bra}[1]{{\langle {#1} |\,}}
\newcommand{\ket}[1]{{\,| {#1} \rangle}}

\def\id{\textrm{id}}

\def\id{{1 \kern-.28em {\rm l}}}

\def\journal#1&#2(#3){\unskip, \sl #1\ \bf #2 \rm(19#3) }
\def\andjournal#1&#2(#3){\sl #1~\bf #2 \rm (19#3) }

\def\ie{{\it i.e.}}
\def\eg{{\it e.g.}}

\def\etc{{\it etc}}

\def\sst{\scriptscriptstyle}

\def\One{{1\hskip -3pt {\rm l}}}

\catcode`\@=11
\def\slash#1{\mathord{\mathpalette\c@ncel{#1}}}
\def\underrel#1\over#2{\mathrel{\mathop{\kern\z@#1}\limits_{#2}}}

\catcode`\@=12

\def\bra#1{\left\langle #1\right|}
\def\ket#1{\left| #1\right\rangle}

\def\exp{{\rm exp}}

\def\ie{{\it i.e.}}
\def\eg{{\it e.g.}}

\title{
Trouble in paradox
}

\author{Emil J. Martinec}

\affiliation{
\vskip 0.01cm
Kadanoff Center for Theoretical Physics and Enrico Fermi Institute\\ 
University of Chicago\\ 
5640 S. Ellis Ave.\\
Chicago IL 60637\\ 
}

\emailAdd{%
e-martinec@uchicago.edu}

\abstract{%
Recent developments in holography have suggested a potential resolution to the black hole information paradox within the context of gravitational effective field theory.  We emphasize the non-local nature of this proposed resolution, and highlight the ways in which observations outside the black hole can detect it and conclude that the black hole is not radiating like an ordinary body would.
}

\begin{document}
\hypersetup{pageanchor=false}
\begin{titlepage}
\maketitle
\thispagestyle{empty}
\end{titlepage}
\hypersetup{pageanchor=true}
\pagenumbering{arabic}





\section{Paradox Lost?} 
\label{sec:intro}

With the advent of microscopic accountings of the density of states of a black hole in string theory~\rcite{Strominger:1996sh}, it seemed that a resolution of the black hole information paradox might finally be at hand.  But while these accountings provided exemplars of unitary theories of quantum gravity, the horizon scale physics remained obscure and so the flaw in the arguments leading to the information paradox remained similarly obscure. 

One suggestion to resolve the information paradox that arose not long after this breakthrough was that of black hole complementarity~\rcite{Susskind:1993if}~-- that the black hole interior was secretly encoded in the Hilbert space of the exterior in a complicated way, such that the interior and exterior observables did not commute.  The latter property was supposed to preclude various inconsistencies which might arise from various thought experiments in which observers measure the Hawking radiation, and then dive into the black hole to measure the state of the interior.  The development of the BFSS matrix model~\rcite{Banks:1996vh} hinted that something like black hole complementarity was on the right track; in this model (one of the first instances of gauge/gravity duality), the fundamental degrees of freedom specifying the locations of objects in spacetime are matrices, which become non-commuting in black hole states.
But the details of how this complementarity was to be implemented in practice were never spelled out. 

Subsequently, a careful examination of the quantum entanglement properties of states in quantum field theory led to a realization that the notion of black hole complementarity could not resolve the information paradox, and to a fuller appreciation of the fundamental incompatibility of unitarity, locality, and causality in the effective field theory description of black hole evaporation~\rcite{Mathur:2009hf,Braunstein:2009my,Almheiri:2012rt}.  This realization has led to a sharpening of the information paradox.

\begin{figure}[ht]
\centering
\includegraphics[width=1.0\textwidth]{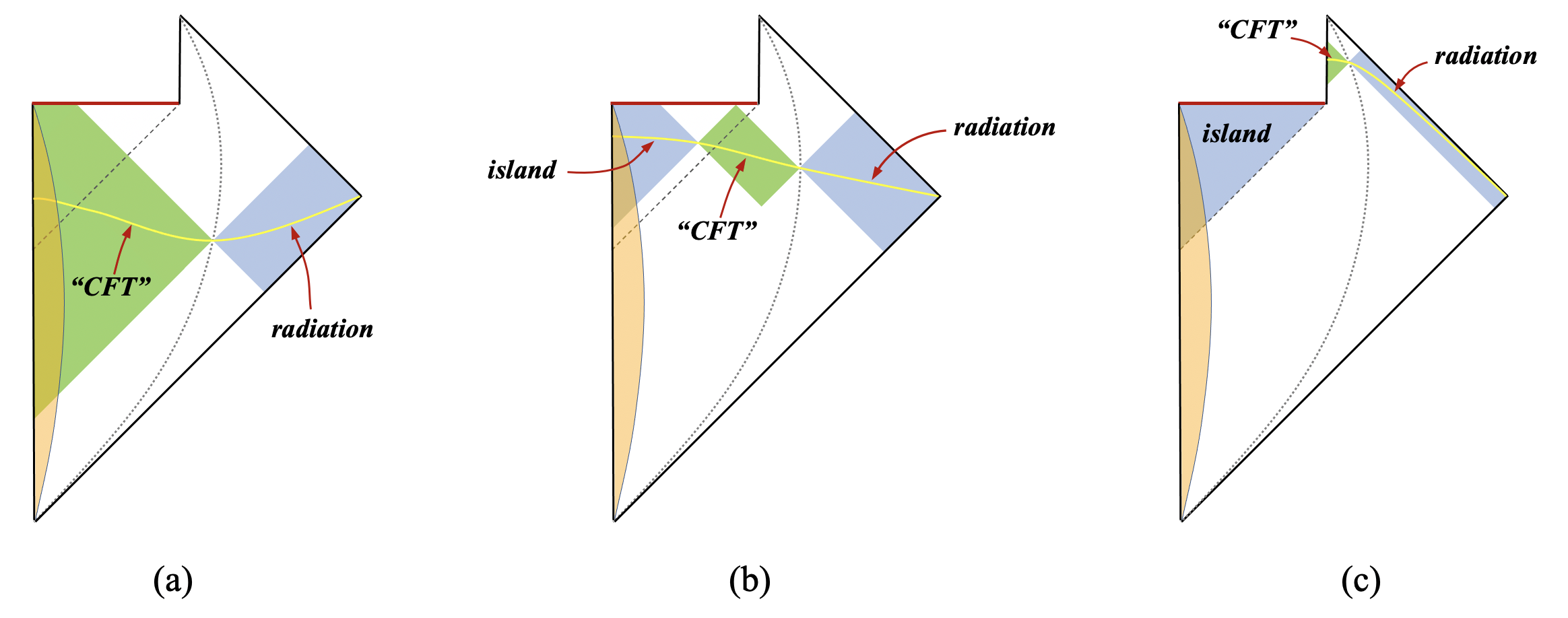}
\caption{\it The standard Penrose diagram for black hole formation and evaporation a holographic CFT coupled to an external radiation bath, showing different Cauchy slices (yellow) in the island scenario.  The matter forming the black hole occupies the region in orange; the entanglement wedge of the CFT is shaded green, while that of the external bath is shaded blue.}
\label{fig:island-scenario}
\end{figure}

Recently, though, a variant of the idea of black hole complementarity has arisen, as a circle of ideas regarding the nature of holography has taken hold.  This circle of ideas has crystallized in the following chain of logic:
\begin{enumerate}[start=1,
    labelindent=\parindent,
    leftmargin =1.7\parindent,
    label={\it(\roman*)}]
\item
Quantum entanglement builds geometry~\rcite{Maldacena:2001kr,Ryu:2006bv,VanRaamsdonk:2010pw}.  
\item
Geometry and entropy are intertwined notions bound together by entanglement, as reflected in the generalized entropy formula associated to a surface $X$ on a Cauchy slice $\Sigma$.  
\be
\label{genent}
S_{\rm gen}(X) = \frac{A(X)}{4G_N} + S_{\rm semi-cl}\big(\Sigma_X\big) ~,
\ee
where $S_{semi-cl}$ is the entropy of effective field theory on the hypersurface $\Sigma$ outside of $X$.  A {\it quantum extremal surface} (QES) extremizes this expression, varying both the slice $\Sigma$ and the surface $X$ subject to appropriate boundary conditions.  In the context of AdS/CFT, an {\it entanglement wedge} is the bulk domain of dependence of spatial hypersurface bounded by a QES $X$ and a homologous domain of the conformal boundary $\cB$~\rcite{Engelhardt:2014gca}. 
\item
The proposed resolution of the black hole information paradox involves the careful application of this formula to the dynamics of evaporating black holes~\rcite{Penington:2019npb,Almheiri:2019psf}.  Early on, the dominant extremizing surface $X$ is trivial (see figure~\ref{fig:island-scenario}a), and the generalized entropy is that of Hawking radiation embodied in the second term of~\eqref{genent}; but after the black hole has evaporate halfway (\ie\ past the {\it Page time}), extremization of the generalized entropy leads to the emergence of a more dominant saddle in which $X$ is a quantum extremal surface near the black hole horizon, as depicted in figure~\ref{fig:island-scenario}b.  
\item
A key component of this proposed resolution is the transfer of the portion of the black hole inside the QES, known as the {\it island}, to a subsystem of the radiation Hilbert space.  The Hawking quanta exterior to $X$ are entangled with and purified by quanta on the island; if the island is part of the radiation, then these entangled pairs are in a pure state in the radiation Hilbert space and will contribute nothing to the generalized entropy, which is now dominated by the area term.  The generalized entropy now decreases as the black hole continues to evaporate; the island grows to encompass more and more of the black hole interior, consistent with unitarity.
After the black hole has evaporated, the black hole interior is entirely contained in the radiation state space, as indicated in figure~\ref{fig:island-scenario}c.
\end{enumerate}
This {\it island scenario} for black hole evaporation is sketched in figure~\ref{fig:island-scenario}.
%
%

There remains however a basic tension between the effective field theory description of the black hole interior and the recovery of information during the evaporation process.  The analysis of~\rcite{Mathur:2009hf,Braunstein:2009my,Almheiri:2012rt} has not been superseded so much as largely ignored, in that the process by which the island region of the black hole interior is transferred to the radiation Hilbert space is left unspecified.  One needs more than entanglement~-- one needs a quantum channel (\ie\ Hamiltonian dynamics) by which the information is transferred to the radiation state space.  Unitary operations restricted to one member of a pair of entangled subsystems will not accomplish this task.
Below we will argue that when the options for such a process are considered more carefully, the information puzzle returns.%
\footnote{Some of our analysis overlaps with that of~\rcite{Almheiri:2013hfa}.}
Any fix which retains the Hawking process at the horizon requires non-local processes that can be detected by observations exterior to the black hole.  Such observers would conclude that the black hole is not radiating like an ordinary body would.


\section{The Hawking process} 
\label{sec:hawking}

The essence of the Hawking process is that a smooth foliation of spacetime near the horizon of a black hole leads to a stretching of the spatial geometry.  This time-dependence continually pulls up modes from the UV and separates them spatially.  The UV vacuum entanglement structure, re-expressed in terms of the stretched modes of the out-state, leads to the (Unruh) state
\be
\ket{ \Psi_{\rm Unruh} } = \prod_\omega\,\exp\big[ e^{-\beta\omega}\, b^\dagger_\omega c^\dagger_\omega\big]  \ket{ \Psi_{\rm Boulware} }
\ee
that is seen by an external observer as a stream of outgoing particles.  Here $\ket{\Psi_{\rm Boulware}}$ is the vacuum for $b$ quanta exterior to the horizon and $c$ quanta interior to it, while $\omega$ is their frequency referred to the asymptotic region, and $\beta$ is the inverse temperature of the black hole.

\begin{figure}[ht]
\centering
    \includegraphics[width=.35\textwidth]{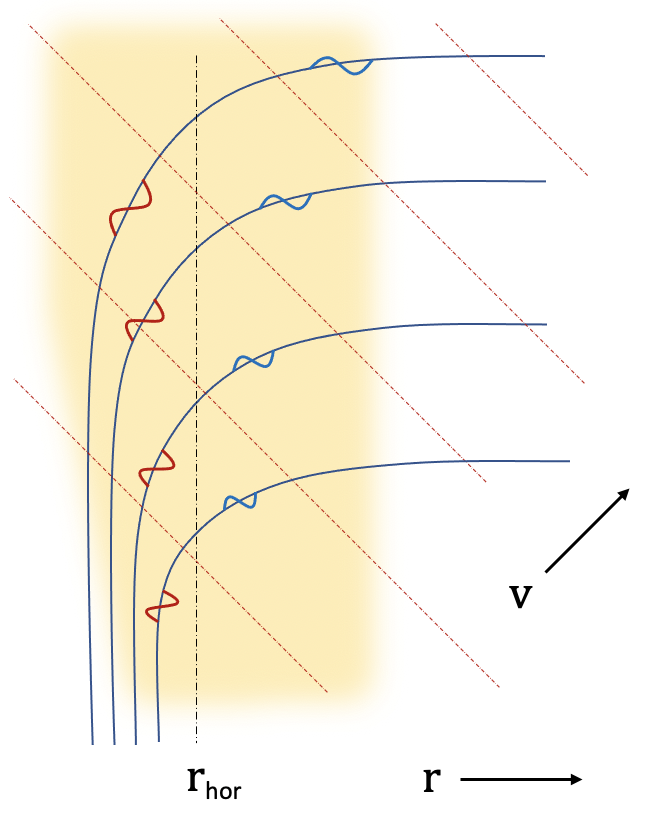}
\caption{\it 
Eddington-Finkelstein radial and ingoing-null coordinates $(r,v)$ near a black hole horizon, displaying a sequence of ``nice slices'', \ie\ smooth spacelike hypersurfaces that avoid the black hole singularity.  The stretching of these hypersurfaces under time evolution can be taken to be largely confined to the orange-shaded region near the horizon.  The time dependent stretching of the geometry in this region in this slicing generates Hawking pairs (red and blue) traveling along outgoing null trajectories.
}
\label{fig:HawkingPair}
\end{figure}

Working in position space rather than frequency/wavenumber space, one can model this state as a sequence of entangled qubits generated in the near-horizon geometry (see \eg~\rcite{Mathur:2009hf}), one of which is carried off to the far region, and the other of which remains in the black hole interior.   In Eddington-Finkelstein coordinates the process appears as in figure~\ref{fig:HawkingPair}.
For a black hole of inverse temperature $\beta$, one can choose a foliation such that the stretching of spatial slices takes place in the vicinity of the horizon (indicated by the orange region in the figure), with the slices stretching an amount of order $\beta$ in a time of order $\beta$, during which on average a single Hawking pair is created.  Thus one can approximate the Hawking process as the creation of a sequence of entangled pairs 
\begin{align}
\label{paircreate}
\ket{\Psi(t_{n+1})} &\approx \ket{\Psi(t_{n})}\otimes \ket{\psi_{\rm pair}}
\nn\\[.2cm]
\ket{\psi_{\rm pair}} &=  \frac 1{\sqrt2}\big( \ket{0}_\sfb\ket{0}_\sfc+\ket{1}_\sfb\ket{1}_\sfc  \big)
\end{align}
such that the state after a modest number $n$ of such creation processes is well approximated as
\be
\label{Psi-at-tn}
\bigl|{\Psi(t_n\sim t_0+n\beta)}\big\rangle \approx 
 \ket{\Psi(t_0)}\otimes \big(\ket{\psi_{\rm pair}}\big)^{\otimes n} ~.
\ee
The process~\eqref{paircreate} is an isometric embedding of the state $\ket{\Psi_\BH(t_{n})}$ at time $t_n$ into a larger Hilbert space containing two additional qubits at time $t_{n+1}$.

We should not take $n$ to be too large, for instance one only needs this picture to be valid over the time it takes to emit a few Hawking quanta, so that the full state of the black hole can then be approximated as the prior state of the black hole $\ket{\Psi_\BH}$ (which we model as a state $\ket{\Psi_\sfA}$ of $M$ qubits that we label by $\sfa$), tensored with the state $\ket{\psi_{\rm pair}}^{\otimes n}$.
We do not allow an order one modification of the state of the radiated $\sfb$ and $\sfc$ qubits over short time scales (\ie\ times much less than the scrambling time), under the assumption that the near-horizon effective field theory applies.  Over larger time scales, we should allow for the rapid scrambling of degrees of freedom in the black hole interior, which we model as some pseudorandom unitary $U_{AC}\in U(2^{M+n})$ acting on the $\sfa$ and $\sfc$ qubits in $\ket{\Psi_\BH}\in \cH_\sfA\otimes\cH_\sfC$
\begin{align}
\label{scramble}
\bigl|{\Psi_\tot(t_0+t_{\rm scr})}\bigr\rangle & \approx  U_{AC} \Big(\ket{\Psi_\BH(t_0)}\otimes \ket{\psi_{\rm pair}}^{\otimes n}  \Big)
\nn\\[.2cm]
& = \sum_{\sigmab_\sfb} \ket{\Psi_\BH(\sigmab_\sfb)}\otimes\ket{\sigmab_\sfb}
\end{align}
(here $\{ \ket{\sigmab_\sfb} \}$ is a basis for $\cH_\sfB$, and $t_{\rm scr}\sim\frac1{2\pi}\beta\log (S\!-\!S_0)$ is the scrambling time)
such that the $\sfb$ quanta are now maximally entangled with the scrambled degrees of freedom in the remaining black hole.  
Whatever the scrambling dynamics is, it should be internal to the black hole and should not itself transfer information to the radiation, for the following reasons.
First, a one-sided unitary such as~\eqref{scramble} applied to two entangled subsystems does not transfer information to the complementary subsystem $\sfB$.
Second, by the point in time where scrambling of~\eqref{Psi-at-tn} becomes important, the $\sfb$ quanta are well away from the black hole, and so any unitary acting to mix them with the $\sfa$ and/or $\sfc$ quanta in the black hole interior amounts to a violation of local effective field theory.

Because the qubit pairs are created in a unique state, the Hawking pair component of the state~\eqref{Psi-at-tn} carries no information.  Scrambling it according to~\eqref{scramble} does not encode any information in $\sfB$.  The state of the radiated $\sfb$ quanta does however have an ever-rising entanglement entropy with the black hole interior, which is the source of Hawking's conclusion that black holes violate unitarity.

The small corrections theorem~\rcite{Mathur:2009hf} and its elaboration in~\rcite{Guo:2021blh} forbids a resolution to this conundrum via small modifications of the evolution due to quantum gravity effects~-- the entropy of the radiation increases by an amount close to $\log(2)$ with each emitted quantum.  For instance, the reduction in the energy of $\ket{\Psi_\BH}$ due to the radiation is a small kinematic effect by which the radiation in the near zone interacts with the black hole interior, which does not modify the radiation process at leading order.

The essence of this theorem is that 
\begin{enumerate}[start=1,
    labelindent=\parindent,
    leftmargin =1.7\parindent,
    label=(SC\arabic*)]
\item
\label{scta}
If a black hole has a smooth horizon where effective field theory is valid, Hawking quanta are pair created in a fully entangled state $\ket{\psi_{\rm pair}}$~\eqref{paircreate}, up to small corrections.  Once created, the pairs are swept apart by the stretching of the near-horizon geometry and soon become well-separated in space.  In the context of local effective field theory, they subsequently do not interact.
\item
\label{sctb}
Due to the monogamy of entanglement and strong sub-additivity of subsystems of the Hawking pairs, the von Neumann entropy of the Hawking radiation state is then monotonically rising.
\end{enumerate}
This result is particularly robust; the subsystems involved in the application of strong sub-additivity are the previously radiated Hawking quanta $\{\sfb\}$, and the next created pair $\sfb_{n\tight+1},\sfc_{n\tight+1}$, and so the analysis does not care about the internal dynamics of the black hole on time scales of order the scrambling time or longer such as~\eqref{scramble}. 
The powerful constraint the (effective) small corrections theorem imposes on any resolution of the information paradox continues to be somewhat underappreciated.

The pair creation process not only does not radiate away information about the initial state of the black hole, it creates a further ``entanglement deficit'' that must be made up for somehow if unitarity is to be maintained.  But the radiation carries away energy in addition to entanglement, and so this deficit has to be repaid, as well as the initial information radiated, with fewer resources than one had in the initial state.  If one waits too long to begin addressing this issue, little energy remains in the black hole state, and one must emit a large amount of entropy using little energy, which takes a long time;%
\footnote{This correlation between the energy used to emit the purifying quanta and the time it takes for them to be emitted can be seen explicitly in moving mirror models where a mirror accelerates away for some period of time and then stops accelerating~\rcite{Wald:2019ygd}.  There is a pool of quanta near the mirror that purify the Hawking-Unruh quanta generated during the accelerating phase that are released once the mirror stops accelerating, but they are exponentially redshifted due to the velocity of the mirror relative to its initial rest frame, and such low-energy quanta take exponentially long to be emitted.} 
one has a black hole remnant with all its attendant inconsistencies (particularly in the context of AdS/CFT, where we think we understand the density of states, and there are no such remnants).

A standard setup for the analysis of black hole evaporation in AdS/CFT is to couple the CFT to an auxiliary system or bath.  The earliest computations of the evaporation process in AdS/CFT~\rcite{Callan:1996dv,Das:1996wn} indeed proceeded by adding a term to the CFT Hamiltonian involving the the operator $\cO_\CFT^\phi$ dual to some bulk scalar field $\phi$.  The coupling of such a bulk mode to a bath is schematically
\be
\label{Hint}
H_{\rm int} =  \lambda_\phi \int \! d^dx\, \cO^\phi_\CFT(x) \, \cO_{\rm bath}(x)  ~.
\ee
This coupling reproduces bulk dynamics of a near-extremal black hole in asymptotically flat spacetime as follows.
If the AdS decoupling limit is not taken, and the AdS throat opens out into asymptotically flat spacetime at some radial scale $r_{\rm throat}\gg r_{\rm hor}$, then the solution to the bulk wave equation proceeds by matching the large radius behavior of wave modes in AdS to an outgoing spherical wave in flat spacetime (see for instance~\rcite{Maldacena:1996ix}).  
The modeling of the asymptotically flat region could for instance consist of field theory in a large volume in which $\cO_{\rm bath}$ is a spherical source of size $r^{~}_{\rm AdS}$.
The holographic map near the AdS boundary
\be
\label{opmap}
\lim_{r\to\infty} r^\Delta \phi^{~}_{\rm bulk}(r,x) = \cO_{\CFT}(x)
\ee
relates the large radius asymptotics of a bulk supergravity field to a corresponding local operator in the CFT (whose conformal dimensions is $\Delta$).
At leading order, one can thus mock up the transmission of an excitation of $\phi$ from the AdS region into the asymptotically flat region in the non-decoupled geometry, via the operator~\eqref{Hint} in the decoupled theory whose effect is to absorb a quantum at the top of the AdS throat and transfer the energy to a quantum in the bath~\rcite{Chowdhury:2007jx,Chowdhury:2008uj,Almheiri:2013hfa}.  
For instance, for a minimally coupled scalar in the full geometry where an $AdS_3\times\bS^3\times \bT^4$ throat opens out into flat spacetime at a radius $r_{\rm throat}$, the decay rate is proportional to~\rcite{Avery:2009tu}
\be
\label{decayrate}
\frac{d\Gamma}{dE} \propto \left(\frac{R_{AdS}^2 E}{r_{\rm throat}}\right)^{2(\ell+1)} 
\left| \bra{{\it final}} \cO^\phi_{\CFT} \ket{{\it init}} \right|^2
\ee
This result tells us that, absent peculiar non-local effects, the CFT operator employed in the interaction Hamiltonian~\eqref{Hint} acts on the $\sfb$ qubits of the Hawking radiation when they reach the AdS boundary and doesn't touch the $\sfc$ qubits, if the Hawking process (with small corrections) is occurring at the horizon.

In principle, we would include such a coupling for all the bulk modes of the effective supergravity theory; however, for addressing issues of principle, we always have the option of tailoring the interaction to include only particular modes of particular fields according to the issue at hand.

We will thus simplify matters in several respects.  First, we can restrict to S-wave modes since these are the dominant component of Hawking radiation; the coupling is then only dependent on time along the AdS boundary.  We take the bath to be a free field in a large volume, with the operator $\cO_{\rm bath}$ a localized source (smeared over a spherical region of size $r_{\rm AdS}$) and the coupling $\lambda_\phi$ controlling the rate at which $\sfb$ quanta reaching the top of the throat are transmitted into the bath as opposed to reflected back down the throat.  
We might for instance turn down the coupling by decreasing~$\lambda_\phi$, to the point that Hawking quanta are transferred to the bath one at a time at well-separated intervals, radiating out from the source region never to return (or at least, not on the time scale it takes for the evaporation to complete).  We can thus treat the bath as a collection of free field modes, and the state generated by the coupling to the CFT as occupying such modes, spatially and temporally well separated and thus non-interacting.%
\footnote{The setup is designed so as to avoid the possibility that the radiation modes interact with one another via their coupling to the CFT at the source.}
Once again, we further pare down the model by treating the bath modes as a bunch of qubits $\sfr$ in a Hilbert space $\cH_\sfR$, initially all in the ``vacuum'' state $\ket{0}_\sfr$.


\section{Paradox regained} 
\label{sec:islands}

In the attempt to rescue unitarity while maintaining effective field theory in the vicinity of the horizon, the island scenario transfers much of the black hole interior to the radiation state space, in order to restore purity to the radiation state after the black hole has evaporated.  
But what could it mean to say that the island ``becomes part of the radiation'', and how does it evade the small corrections theorem and the resulting entanglement deficit?

A key ingredient of the small corrections theorem is bulk locality; once created, Hawking pairs separate and the radiated member of the pair no longer interacts with the black hole.  The theorem assumes that small corrections can only happen while the nascent Hawking quanta are in the vicinity of the horizon, and that once the exterior member of the pair is well-separated from the black hole the interactions vanish sufficiently rapidly that they can be ignored.

\vskip .8cm
\noindent
\ref{sec:islands}.1~~{\em Interactions between the black hole and Hawking quanta within AdS}
\medskip

The conventional black hole evaporation process in the model of section~\ref{sec:hawking} was reduced to evolution under the interaction Hamiltonian $H_{\rm int}$~\eqref{Hint} which implements a unitary transformation 
\be
\label{expHint}
U_{B,R} = e^{iH_{\rm int}\,\delta t} 
\ee
mixing $\cH_\sfB$ with $\cH_\sfR$, extracting $\sfb$ qubits from $\cH_\sfB$ and depositing them in $\cH_\sfR$.
If this were all there is, and there were no non-local interactions of the sort~\eqref{nonlocal}, one would have simply the Hawking process and a continually rising entropy in the bath subspace $\cH_\sfR$, as dictated by the small corrections theorem.

Non-local interactions between the radiated quanta and the black hole bypass the assumptions of the theorem and could thereby evade its conclusion of monotonically rising entropy.  For instance, in addition to the scrambling dynamics of the internal degrees of freedom of the black hole according to~\eqref{scramble}, a small correction to the evolution could also modify slightly the state of the radiated $\sfb$ quanta
\be
\label{nonlocal}
\ket{\Psi_{\tot}(t_{n+1})} = U^\epsilon_{AC,B}\Big( U_{A,C}\ket{\Psi_\BH(t_n)}\otimes \ket{\psi_{\rm pair}} \Big) ~.
\ee
Here $U_{A,C}$ is a scrambling dynamics internal to the black hole and $U^\epsilon_{AC,B}$ is an infinitesimal unitary rotation in the full CFT Hilbert space that affects the radiated $\sfb$ quanta in a way that depends on the internal state of the black hole.  The accumulated effect after many time steps is the sort of large correction needed to evade the small corrections theorem.  For instance, this infinitesimal transformation can partially swap the $\sfa$ and $\sfb$ qubits, so that by the time the $\sfb$ quanta have reached the AdS boundary they have been fully swapped out for the original $\sfa$ qubits, which can then be extracted into the bath.  At this point those $\sfb$ qubits have been embedded in $\cH_\BH=\cH_\sfA\otimes\cH_\sfC$ together with whatever their $\sfc$ partners have evolved into, and the Hawking pair that was introduced by the vacuum dynamics near the horizon can be eliminated via the scrambling dynamics of the black hole.

In this scenario, there are nonlocal interactions prior to the Hawking quanta reaching the AdS boundary, such that the interaction~\eqref{expHint} at the boundary is instead radiating the initial $\sfa$ qubits into the radiation state space.  
The supposed dynamics acts on the radiated Hawking quanta to change their state far from the black hole (but while they are still in the AdS region), in such a way that there are no additional radiated quanta, but the interaction gradually restores purity to the radiation state as the system evolves.  
To summarize the scenario, after Hawking pairs are created by the standard effective field theory mechanism sketched above, the system executes a swap operation that replaces the $\sfb$ qubits with the $\sfa$ qubits as they propagate out, and thereby allows the $\sfb$ quanta to recombine with their partners $\sfc$ and disappear along with the black hole in a unitary fashion while the $\sfa$ quanta comprise the radiation in the end.

This dynamics is inherently non-local, connecting the black hole interior to arbitrarily distant regions.  Can such wild non-locality be detected by an observer remaining outside the black hole, or is it somehow hidden in the complexity of the encoding of bulk physics in holographic systems?

A basic difficulty with such a scenario is that we are free to non-destructively measure the Hawking quanta $\sfb$, and ask whether their state is changing as they propagate outward.
Suppose we erect a set of ``quantum non-demolition'' measuring devices, that don't change the bit-parity of the $\sfb,\sfr$ qubits but correlate them with a measuring apparatus $\sfM$ with Hilbert space $\cH_\sfM$.  We begin with $k$ measuring apparatus qubits $\sfm$ in a unique state, say $\ket{\Psi_\sfM}=\ket{0_1 0_2\ldots 0_k}_\sfm$, and let the measurement consist of the encoding
\begin{align}
\ket{0}_\sfb\otimes \ket{\Psi_\sfM} &\longmapsto \ket{0}_\sfb\otimes \ket{0_1 0_2\ldots 0_k}_\sfm
\nn\\[.1cm]
\ket{1}_\sfb\otimes \ket{\Psi_\sfM} &\longmapsto \ket{1}_\sfb\otimes \ket{1_1 1_2\ldots 1_k}_\sfm 
\end{align}
which is a simple example of an error-correcting code.
We can arrange to have a series of such measuring devices placed at various radial positions in order to record the state of a Hawking quantum as it propagates out toward the AdS boundary.  Afterward, we can read out the states of the measuring devices and determine whether the qubit state has flipped during the course of its evolution.  If the outgoing radition were interacting with the black hole interior in some non-local way and swapping $\sfb$ qubits for $\sfa$ qubits, then the outgoing qubit has a substantial likelihood (of order 50\%) of being flipped along the way.  We can similarly examine whether the qubits in $\cH_\sfB$ are faithfully transferred to bath quanta $\sfr$, by measuring the state in $\cH_\sfB$ before and  comparing it to the state in $\cH_\sfR$ after.%
\footnote{The conspiracy-minded might want to invoke here the possibility that since the measuring apparatus is also built out of CFT degrees of freedom, the non-local interactions might also reach into it and mess with its internal state as well, thereby erasing our ability to tell whether the Hawking qubit has flipped during propagation.  But the non-local interactions would then have to know about the details of the measuring apparatus (which might consist of a large number of qubits), and whether we have sent a signal from the boundary to turn it on or off, \etc.; and such a possibility becomes ever more preposterous.}

In this way, we can detect whether the black hole is radiating quanta as an ordinary thermal body would radiate (\ie\ whether the state of the radiated quanta decouples from that of the black hole once it leaves the vicinity of the horizon), or if instead there are non-local effects persisting in faraway regions.

Of course, it could happen that this swap of the $\sfa$ and $\sfb$ qubits takes place in a small region or ``atmosphere'' of the black hole, so that a few horizon radii out the black hole is radiating like an ordinary body.  But the Hawking pair state $\ket{\psi_{\rm pair}}$ is created in a region of size $r_{\rm hor}$ consisting of quanta whose wavelength is of order $r_{\rm hor}$.  If the Hawking quanta have been repatriated to the black hole interior in just a few horizon radii, essentially one has admitted that local effective field theory breaks down at the horizon scale, since there are large corrections to the Hawking process whereby quantum information is being released and Hawking quanta re-absorbed within just a few horizon radii.%
\footnote{Note that a resolution of this sort would have little to do with the island scenario; the black hole interior is not becoming part of the bath $\sfR$ so much as communicating directly with the near zone exterior $\sfB$.}

Additional arguments against such a scenario were given in~\rcite{Almheiri:2013hfa}.  For instance, one can perform operations on the outgoing radiation that interfere with the swap operation between $\sfa$ and $\sfb$ qubits, thereby decreasing the outflow of information from the black hole, leading to a buildup of entropy that leads to a remnant problem.

\vskip .8cm
\noindent
\ref{sec:islands}.2~~{\em Nonlocal interactions that radiate additional quanta}
\medskip

Another possibility one might entertain is the radiation of additional quanta, apart from those generated by the Hawking process (see for instance%
~\rcite{Giddings:2009ae,Giddings:2011ks,Giddings:2012gc,Giddings:2013kcj,Giddings:2013noa,Giddings:2014nla,Giddings:2017mym,Almheiri:2013hfa,Wald:2019ygd}), 
and that these additional quanta are responsible for emitting the information content in the $\sfa$ qubits of the initial state $\ket{\Psi_\BH}=\ket{\Psi_\sfA}$, as well as the $\sfc$ quanta that purify the state of the $\sfb$ quanta.

The island scenario proposes an effect which is a unitary rotation 
\be
\label{islandformation}
\ket{\Psi_\tot(t_{n+1})} = U_{AC,R} \, U_{B,R} \Big( \ket{\Psi_\BH(t_n)}\otimes\ket{\Psi_\sfB(t_n)}\otimes\ket{\Psi_\sfR(t_n)} \Big)
\otimes \ket{\psi_{\rm pair}}
\ee
mixing the black hole interior state space $\cH_\BH=\cH_\sfA\otimes\cH_\sfC$ with $\cH_\sfR$, so that the earlier qubits $\sfc$ and the qubits $\sfa$ in the black hole interior eventually become ``part of the radiation'', while the Hawking process~\eqref{paircreate} continues at the horizon, and the unitary rotation~\eqref{expHint} transfers $\sfb$ quanta to the bath.%
\footnote{The possibility that $U_{AC,R}$ and $U_{B,R}$ interfere with one another, and modify the transfer of $\sfb$ quanta into the bath, is a variant of~\eqref{nonlocal} where the non-local interaction happens at the very top of the throat where it connects to the bath, and suffers the same problems.}

Note that in the way we have set up the problem, we are only coupling the CFT and the bath through a coupling of the sort~\eqref{Hint}, where $\sfR$ is a free field theory and $\cO_{\rm bath}$ is a localized source.  After $\sfr$ quanta are generated in the bath, they completely decouple from the CFT (since we can make the bath as big as we like), and so any non-local interaction generating~\eqref{islandformation} does not act on these faraway qubits in the bath.  Thus, if the state of the Hawking $\sfb$ quanta has only undergone small corrections, and is transmitted faithfully to the bath via~\eqref{expHint}, then the only way to restore purity to the final state is through the radiation of additional quanta via the same interaction~\eqref{Hint}, having the effect of $U_{AC,R}$ in~\eqref{islandformation}.

\begin{figure}[ht]
\centering
    \includegraphics[width=.35\textwidth]{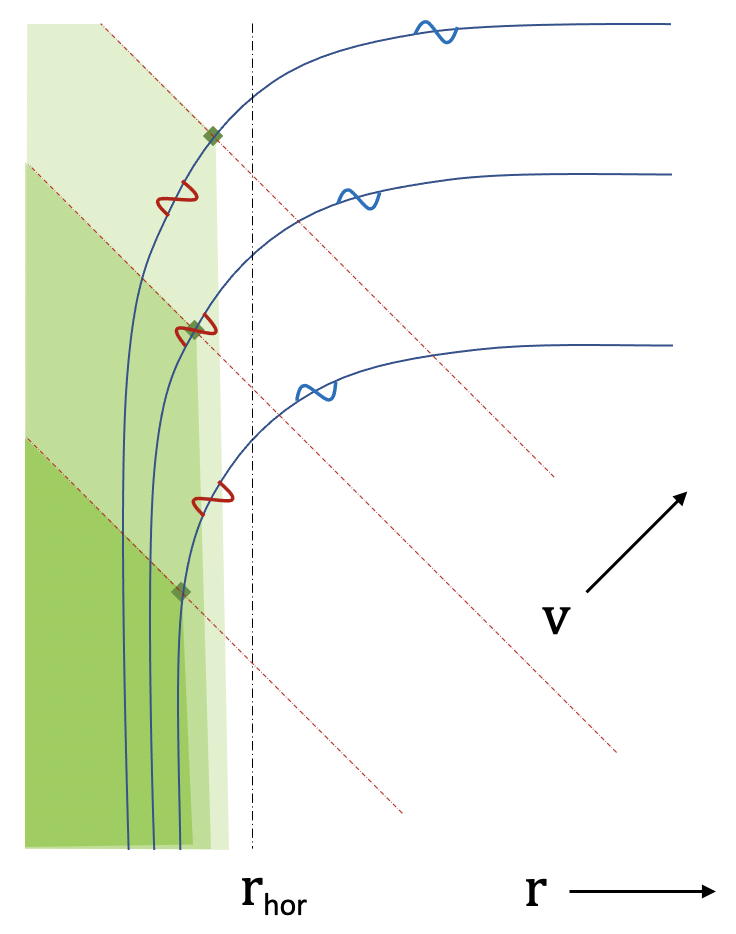}
\caption{\it 
The partner $\sfc$ (red) of a Hawking quantum $\sfb$ (blue) is transferred to the island.  As time evolves, these quanta both travel outgoing null trajectories while the island (shaded green) moves closer to the horizon.  The QES on each time slice is indicated by a green diamond.  In the earliest time slice, the $\sfc$ quantum is in the entanglement wedge exterior to the QES.  The middle time slice depicts the moment of the transition.  In the latest time slice, the $\sfc$ quantum lies in the island.
}
\label{fig:island-xfer}
\end{figure}

If we take the evolution of figure~\ref{fig:island-scenario} literally, there is a non-local process by which $\sfc$ qubits are transferred from the entanglement wedge of the CFT into the island, which is supposed to be a subspace of the radiation state space $\cH_\sfR$.  If we focus on the late-time evolution well after the Page time, the QES gradually moves out toward the horizon and so gradually encodes more and more of the $\sfc$ quanta; figure~\ref{fig:island-xfer} depicts the transfer of a Hawking partner qubit $\sfc$ to the island.

Note that in our analysis we always adopt a convention where $\cH_\sfA\otimes\cH_\sfC$ is a subsystem of the CFT Hilbert space, and ask that the appearance of the island as part of the radiation state space be made explicit as an operation that transfers the data in the black hole interior from the CFT Hilbert space into the radiation Hilbert space $\cH_\sfR$ as some sort of unitary encoding, while the ground state in the bath rotates into a unique state in the black hole state space $\cH_\BH=\cH_\sfA\otimes\cH_\sfC$ as it evaporates away.  
For instance the process depicted in figure~\ref{fig:island-xfer} transferring a $\sfc$ qubit to the island (perhaps after some scrambling has taken place within $\cH_\BH$) involves just such a rotation~\eqref{islandformation}.  In the model of section~\ref{sec:hawking}, this rotation has to be implemented by the interaction Hamiltonian~\eqref{Hint}, and seems to suggest a substantial modification of the operator map~\eqref{opmap}.

More generally, one might allow for the possibility that evolution involves a unitary operator $U_{AC,BR}$ that rotates $\cH_\sfA\otimes\cH_\sfC$ into $\cH_\sfB\otimes\cH_\sfR$, so that some of the information is transferred out while the Hawking quanta are propagating from the near-horizon region out to the AdS boundary as discussed above, while the rest happens at the interface between the CFT and the bath through the coupling~\eqref{Hint}.  

This modification does not substantially change the analysis.
A generic transformation of this sort will result in the random appearance of new radiation quanta far from the black hole, which again would be detectable to outside observers who see the flux growing larger further away from the black hole.  In order for this not to happen, the additional flux should come entirely from the region near the black hole.  This is the sort of scenario envisioned in~\rcite{Giddings:2012gc,Giddings:2013kcj,Giddings:2013noa,Giddings:2014nla,Giddings:2017mym}, in which the black hole appears to outside observers to be radiating in a dual mode, via the Hawking process {\it and} via a non-local communication of the interior to the near zone.  This possibility dramatically affects the thermodynamics, as discussed in~\rcite{Giddings:2012dh,Giddings:2013vda}, in ways that disagree with the known thermodynamics of the dual CFT (for instance in $AdS_3$ where the thermodynamics is dictated by conformal invariance).

Regardless, the process depicted in figure~\ref{fig:island-xfer} is quite different~-- a direct radiation of qubits from the interior of the black hole into the bath, which would be seen as quanta emerging into the asymptotically flat region from the top of the AdS throat that were not present lower down in the throat, outside the horizon.  This outcome obviously differs from the process of radiation from an ordinary body.

In an evolution of the sort~\eqref{islandformation}, in which there is direct radiation of information from the black hole interior into the bath, one way to have such a process push out more information might be to make use of higher angular momentum channels~\rcite{Almheiri:2013hfa}.  The Hawking flux is suppressed due to an angular momentum barrier (see \eg~\rcite{Cvetic:1997uw,Klebanov:1997cx} and equation~\eqref{decayrate}), but if the black hole interior connected directly to the far region, there need not be such a suppression of higher angular momentum modes.  One would essentially be arguing that black hole interior radiates information from an angular sphere of size $R_{AdS}$, possibly as far away as the neck between the top of the AdS throat and the asymptotically flat region, with different locations on the sphere radiating independently.
Such radiation might occur in the standard modeling of a CFT/bath interaction via a local coupling at the neck such as that of  $H_{\rm int}$ of~\eqref{Hint}.%
\footnote{But again, the disparity between the Hawking flux coming up the throat and the flux emerging into the asymptotically flat region would be detectable to observers outside the black hole.}
However, nothing requires us to couple the two systems this way; indeed, we could simply couple the bath to the s-wave mode of the scalar $\phi$ that comprises the dominant component of Hawking radiation by spatially averaging the CFT operator $\cO^\phi_{\rm CFT}$ and coupling it to some bath operator of fixed spatial dependence (some sort of antenna).  Then the Hawking modes would leak out, but not so much the information from the black hole interior, and one is then forced into a remnant scenario of the sort we discuss below.  Unless the non-local radiation and the Hawking radiation are emitting via the same channels in more or less the same way, one can engineer a coupling that allows the latter to radiate into the bath but not the former, leading one back into the paradox.

Moreover, the sort of interaction~\eqref{Hint} coupling the CFT to the bath, which models the nexus between the AdS throat and an asymptotically flat region, doesn't seem to generate any additional flux~-- its role seems to be to transfer Hawking $\sfb$ quanta near the AdS boundary to the bath modes which model the asymptotically flat region.  One would be saying that this same interaction can reach into the black hole interior and non-locally extract $\sfa$ and $\sfc$ qubits, and that the correlation functions of the CFT operators involved are decidedly non-thermal, which seems entirely at odds with what we have learned about the AdS/CFT dictionary, where operators such as $\cO^\phi_{\rm CFT}$ in~\eqref{Hint} have thermal correlation functions at a unique temperature~-- that of the Hawking radiation.


\vskip .8cm
\noindent
\ref{sec:islands}.3~~{\em Remnant dangers}
\medskip

One difficulty with the emergence of $\sfa$ and $\sfc$ qubits from the interior was mentioned above~-- that one is now trying to repair the entanglement deficit, as well as radiate away the information in the initial state, with fewer resources.  Furthermore, this additional radiation only adds to the Hawking flux; such a black hole is ``hotter'' than black hole thermodynamics would predict, with a distribution of energies that will be at odds with that of a thermal body.%
\footnote{Of course, the Hawking radiation from a black hole is not a pure blackbody spectrum due to the greybody filter that results from the propagation of the radiation through the ambient geometry to the far region~\rcite{Maldacena:1996ix,Klebanov:1997cx,Cvetic:1997uw}.  But these are determined by the ambient geometry and so in principle one can measure that geometry, compute the expected Hawking flux, and compare to the actual flux.}
Black hole properties such as the equation of state and the radiated spectrum would thus differ from the predictions of semi-classical gravity by a significant amount~\rcite{Giddings:2012dh,Giddings:2013vda}.

Emitting the quanta necessary to maintain the purity of the final state, with the energy available after one devotes a large fraction of the energy budget to the emission of Hawking modes that carry away no information, runs the risk of generating a remnant scenario with all the attendant inconsistencies in an AdS/CFT context (since the CFT density of states is rather well understood, and inconsistent with a large component comprised of remnants).

The system might try to hide this additional flux in extremely soft radiation modes~-- some sort of ``soft hair'' (as for instance in~\rcite{Hawking:2016msc}) that can carry away information at very little cost in energy, and is so diffuse that it is hard to measure.  However such soft hair takes a very long time to radiate a large entropy.  

For instance the most efficient entropy flux for a given energy flux is that of thermal radiation; the entropy flux is bounded by the available energy and so the rate of information emission becomes lower and lower the smaller the fraction of the energy budget of the black hole is devoted to repairing the entanglement deficit and actually radiating the initial information content of the black hole (see~\rcite{Wald:2019ygd} for a recent discussion in the context of moving mirrors). 
If one really wants to radiate the information in soft hair, one is pushing it out via the same sort of field theory modes that carry the Hawking radiation, just at a lower temperature.  
But then the rate of entropy radiation in such modes is much lower, and unitary evolution generates a remnant as one waits for the black hole to recover from the entanglement deficit built up by the Hawking process.

It would seem that one needs the $\sfa$ and $\sfc$ qubits to be emitted at roughly the same rate as the $\sfb$ qubits in order to avoid a remnant problem and a contradiction with the known CFT density of states.  But then the temperature of their component of the radiation will be approximately the same as that of the ordinary Hawking process, since in the setup above, the bath and the CFT only interact via \eqref{Hint} and so both sets of qubits are being fed out through the same interaction Hamiltonian.  But then the difference between the expected Hawking flux and the actual flux should then be obvious~\rcite{Giddings:2012dh,Giddings:2013vda}, and one would find that the black hole does not actually satisfy the predictions of general relativity.


\vskip .8cm
\noindent
\ref{sec:islands}.4~~{\em Non-local modifications of Hilbert space structure}
\medskip

%
Another place the necessary non-locality could lurk is in the Hilbert space structure rather than the Hamiltonian.  It has been suggested that the interior degrees of freedom of the black hole are a complicated, highly nonlocal rewriting of degrees of freedom in the black hole exterior~\rcite{Papadodimas:2013wnh,Papadodimas:2013jku,Susskind:2013tg,Maldacena:2013xja,Verlinde:2013qya}, so that the assumption made in section~\ref{sec:hawking} of an independent Hilbert space $\cH_\sfC$ for interior partners does not hold.
This idea had its own set of issues in the context in which it was originally proposed, see for instance~\rcite{Harlow:2014yoa,Harlow:2014yka,Marolf:2015dia}, though see also~\rcite{Papadodimas:2015jra,Papadodimas:2017qit,Kim:2020cds}.
Note that these discussions in large part don't concern bulk dynamics or how the bulk Hawking process occurs, but rather propose encodings of operators and/or qubits, as properties of the holographic map.  We will have more to say about this in the next section.


The island scenario might be regarded as a newer variant of this idea, in which only the portion of the black hole interior inside the QES is mapped in a complicated, non-local way to a subspace of the radiation Hilbert space $\cH_\sfR$.  One might then entertain the notion that the late interior $\sfc$ quanta don't lie in a Hilbert space $\cH_\sfC$ distinct from this subspace of $\cH_\sfR$, and in fact are some complicated rewriting of the early $\sfb$ quanta.  Since the late $\sfb$ quanta are entangled with the late $\sfc$ quanta, they would then be secretly entangled with the early $\sfb$ quanta which by then have been deposited in $\cH_\sfR$.  One would then conclude that the early and late Hawking quanta are in fact entangled, such that the entanglement entropy of the radiation is decreasing after the Page time.  The effective small corrections theorem is thus evaded, and the problem of increasing entanglement is seemingly solved, due to a lack of factorization of the Hilbert space into interior and exterior subspaces.  

However, one can't have simultaneously that (i) the late $\sfc$ qubits are made out of the early $\sfb$ qubits; (ii) the early $\sfb$ qubits are fully entangled with the early $\sfc$ qubits (up to small corrections); (iii) the late $\sfb$ qubits are fully entangled with the late $\sfc$ qubits (again up to small corrections); without having (iv) that the late $\sfb$ qubits are made out of the early $\sfc$ qubits.  At least, if we are supposing that throughout the evolution the Hawking process is taking place at the horizon, so that we have (ii) and (iii).

At late times, the early $\sfb$ and $\sfc$ qubits are in the bath according to the island scenario, completely decoupled from the CFT~-- the early $\sfb$ qubit because it was directly radiated into the bath through the boundary coupling, and the early $\sfc$ qubit is also supposed to have arrived in the bath as part of the island.  So one would be saying that the late-time Hawking radiation process at the horizon is being carried out in free field theory in the bath, which seems absurd.

Furthermore, the $\sfb$ quanta are always in the entanglement wedge of the CFT, until they are dumped into the bath (for a single-sided black hole, where the QES is inside the horizon), so according to the island formalism, at the time of Hawking pair creation, the late $\sfb$ quanta cannot be made from early $\sfc$ quanta since the latter are supposed to be in the bath already, and quanta cannot simultaneously be in the CFT and in the bath.  Ultimately, the proposal runs into the problem that the Hilbert space {\it is} factorized between the CFT and the bath, and so one can't implement the sort of identification of degrees of freedom necessary to make $\cH_\sfC\subset\cH_\sfR$ for the late-time Hawking process.

Finally, this whole discussion makes no provision for the radiation of the $\sfa$ qubits, the actual information in the black hole.  The Hawking process itself isn't carrying away the original information in the state of the black hole, and the whole discussion seems to be an elaborate mechanism for repaying the entanglement deficit without actually radiating any information.  So the radiation of these qubits would constitute additional quanta beyond those of the Hawking process, with the attendant problems discussed above.


\vskip .8cm
\noindent
\ref{sec:islands}.5~~{\em Something else}
\medskip

Perhaps there is some other mechanism that is being invoked in the statement that the black hole interior ``becomes part of the radiation''; if so, it would be helpful to know what it is.  If the non-locality needed to evade the small corrections theorem lies in the Hamiltonian, the types of unitary evolution discussed above appear to cover the various options, and the conclusion seems to be that if the Hawking process of effective field theory is taking place at the black hole horizon, then one can detect from external observations the non-local processes required to override the random, incoherent nature of the Hawking process in order to restore purity of the final state.  Introducing non-locality in the Hilbert space structure has its own set of issues.



\section{But what about \texorpdfstring{$\ldots$}{} ?} 
\label{sec:whatabout}

A potential way to evade the constraints of the small corrections theorem is through the appearance of non-local interactions, such as those discussed above, which can communicate information in the black hole interior to the outside world.  We have argued that such non-local interactions are problematic.  We now come to the question of whether various other non-localities that have been discussed in the literature are up to the task.

Often the complicated nature of the holographic map between the bulk gravity and non-gravitational CFT sides of AdS/CFT duality is invoked as an essential ingredient of the resolution of the information paradox.  In the island scenario, this complexity is thought to provide a means of hiding the island in plain sight in the radiation Hilbert space~\rcite{Kim:2020cds}.  But does this complication somehow bypass the effective small corrections theorem \ref{scta}-\ref{sctb} of~\rcite{Guo:2021blh}?

Non-local effects that have been considered include the following:
\begin{enumerate}[start=1,
    labelindent=\parindent,
    leftmargin =1.7\parindent,
    label=(NL\arabic*)]
\item
\label{code1}
The holographic map is a non-local encoding of bulk gravitational physics in a dual non-gravitational description.
A qubit localized in the AdS bulk is encoded non-locally over many qubits in the dual CFT, in much the same way that quantum information is stored non-locally in an error-correcting code.  Does this form of non-locality allow us to evade the small corrections theorem?
\item
\label{code2}
This encoding is also believed to be ``state dependent'', \ie\ the embedding of bulk effective field theory quanta into the CFT Hilbert space should depend on the microscopic details of the black hole microstate in a complicated way.  Can this secretly embed information about the interior state of the black hole in the radiation?
\item
\label{code3}
Topology-changing processes can also be non-local phenomena that might create pathways (such as traversible wormholes~\rcite{Gao:2016bin,Maldacena:2018lmt}) for information to leak out of black holes.
\item
\label{code4}
There is the possibility that smallness of matrix elements is compensated by the exponentially large number of states available to the black hole. 
\item
\label{code5}
The non-factorizablility of the Hilbert space of quantum field theory and of gauge/gravitational systems is also sometimes pointed to as a mechanism for resolving the paradox (see for instance~\rcite{Raju:2020smc,Raju:2021lwh} for a recent discussion), as are non-locality/non-factorizability of gravitational dressing.
\item
\label{code6}
Yet another possibility is that the holographic map is only approximate.  Semiclassical bulk configurations are coherent states that might be dramatically overcomplete in the state space of the black hole, as suggested for instance by the rapid growth of interior volume on nice slicings such as those of figure~\ref{fig:HawkingPair}.   There might then be a non-local breakdown of effective field theory when the information capacity of the naive state space (given by the volume of the semiclassical bulk phase space) exceeds the size of the actual black hole interior state space of the exact theory.
\end{enumerate}
Let us consider each of these points in turn.

First of all, it is important not to conflate the two sides of the duality, nor to conflate the map between the two sides with the properties of either one.  Let us assume for the moment that AdS/CFT is a true duality~-- that both sides of the duality admit an exact description, and holography is a statement of the equivalence of these two exact descriptions.  The map between these two descriptions can be very complicated, with localized qubits in the AdS bulk mapping to logical qubits non-locally encoded in the CFT.  But this is a non-locality of the {\it map}, not a non-locality or acausality of the dynamics {\it within} either side of the duality. 

The fact that the holographic map embeds a bulk qubit by distributing it over many qubits in the CFT dual does not alter the fact that in the bulk description there is a $\sfb$ qubit that is entangled with the black hole interior via the Hawking process which, absent nonlocal and acausal effects {\it in the bulk description}, will not evolve in a way that decreases the von Neumann entropy of the black hole exterior subsystem, once the Hawking quantum leaves the vicinity of the black hole.  These logical qubits may indeed be spread over many qubits in the exact CFT description, and some of them may be subject to fast scrambling dynamics that mixes them in some complicated way; nevertheless, if the holographic map is exact, and the Hawking process is hypothesized to be taking place at the horizon in the bulk description, then there will be an image under the map of the Hawking process in which there will be a logical $\sfb$ qubit fully entangled with a logical $\sfc$ qubit in the CFT at each step of the evaporation process.  There will also be the images of the previously emitted logical $\sfb$ qubits, and these plus the next pair are all one needs to apply the small corrections theorem.  Locality of the bulk dynamics dictates that soon after the $\sfb$ qubits leave the vicinity of the black hole, they decouple from the interior dynamics of the black hole, and the degrees of freedom in the CFT that represent it under the duality map.  Both descriptions will have to obey the effective small corrections theorem for these logical $\sfb$ qubits, together with the next newly minted $\sfb,\sfc$ pair.  

One is then forced into a situation where, if there is some CFT-plus-bath dynamics that purifies these logical $\sfb$ qubits, then running the holographic map the other way, it must be bulk-nonlocal.  
Either the logical $\sfb$ qubits are bulk non-locally communicating with the black hole interior in a way that detectably changes their state as discussed in section~\ref{sec:islands}.1, or the boundary interaction~\eqref{Hint} magically acts to both transfer Hawking $\sfb$ qubits to the bath as well as ``island'' qubits $\sfc$ and $\sfa$ that are interior to the black hole, as discussed in section~\ref{sec:islands}.2.  But this latter possibility will also be easily detectable in the bulk as a process by which the flux coming up the AdS throat differs dramatically from the flux entering the bath, and black hole thermodynamics is significantly modified from the Hawking predictions.  If the holographic map is exact, then the bulk dynamics has a boundary image, and the small corrections theorem has a boundary image.  All that the holographic map has done is to obfuscate the bulk dynamics via the complexity of that map.

The supposed state dependence of the encoding map is also a statement about the holographic map rather than the dynamics within a given side of the duality.  It does not change the entanglement structure of the Hawking pairs, and their von Neumann entropy within the bulk description up to small corrections.  Radiation of the information in the initial state of course requires state dependence, but the small corrections theorem tells us that state-dependence of the holographic map does not help.  One needs the state of the radiation to depend on the state of the black hole, and this is what is ruled out by the small corrections theorem absent nonlocal processes {\it in the bulk}, if the Hawking process is occurring at the horizon.

We conclude that \ref{code1} and \ref{code2} are irrelevant to the problem at hand.

Topology-changing processes are an example of bulk non-local effects \ref{code3} that might transfer information out of the black hole interior~-- a particular mechanism for implementing the sorts of non-local dynamics discussed above.  If they act so as to radiate additional quanta beyond those of the Hawking process, then this will affect the distribution of quanta in the radiation in detectable ways, as discussed above~-- the spectrum of quanta radiated into the bath differs significantly from the Hawking prediction.  The possibility that the topology-changing processes don't affect the spectrum but rather gradually imprint the data of the initial black hole state onto the radiation was also discussed above; we saw that there are simple ways to tell if the states of the $\sfb$ qubits were swapped out for the $\sfa$ qubits on their way out to the AdS boundary.
Either way, an outside observer measuring the radiation will observe that the black hole is not radiating as an ordinary body.

Next, we come to the question \ref{code4} of whether the smallness of any individual correction to the semiclassical bulk picture is overwhelmed by the enormous number of black hole microstates.  The problem is that this enormous number has to do with the internal structure of the black hole microstates, whereas the small corrections theorem involves the application of strong sub-additivity to subspaces of the radiated quanta that are far from the black hole and seemingly decoupled from the black hole interior.  Any correction in which the large number of interior states of the black hole affects the radiated quanta amounts to a non-local effect of the sort discussed above~-- an order one mixing between the Hawking radiation and the black hole interior, that could be detected by outside observers making non-destructive measurements on the radiation.

Regarding the non-factorizability of the Hilbert space \ref{code5}, while this is a feature of the ultraviolet structure of field theory it should not be an issue for the effective dynamics as expressed \eg\ in bit-models of the evaporation process.  The fact that the fields whose quanta are electrons or photons have a highly entangled ultraviolet vacuum structure, as well as a long-range gravitational (and for the electron, electromagnetic) dressing, does not prevent us from isolating the spin states of a few electrons in a magnetic trap or photons in a superconducting cavity and approximating their dynamics to a high degree of accuracy using a factorized Hilbert space description.  These are the sort of small effects that are accommodated by the effective small corrections theorem.  
Exotic forms of non-factorizability that might affect the analysis of effective bit-models of the evaporation process were dealt with in section~\ref{sec:islands}.4.

Corrections due to gravitational dressing are non-local and spoil factorizability.
However, such effects are purely kinematic in nature, and in 3+1d have been argued to be exponentially small at least in perturbation theory~\rcite{Donnelly:2018nbv,Giddings:2021khn}.  Basically, these tails of the wavefunctions of physical states carry very little information beyond global charges.  Multipole moments of the fields that might distinguish microstates are highly suppressed in generic states (both above the black hole transition where the observables have to approximate the classical no-hair theorems, and at threshold where \eg\ the BPS spectrum and its properties admit quantitative analysis~\rcite{Balasubramanian:2005qu,Balasubramanian:2018yjq}).

In 1+1d, gravitational dressing is well understood.  
In two-dimensional dilaton-gravity models, the gravity sector consists of the scale factor $\rho$ of the metric (so that the dynamical metric $g_{\alpha\beta}=e^{2\rho}\hat g_{\alpha\beta}$ is a Weyl rescaling of a fixed metric $\hat g_{\alpha\beta}$) together with a dilaton $\phi$.  These can be traded for a pair of ``light-cone coordinates'' $X^\pm$ in field space via a canonical transformation.
Two-dimensional spacetime dynamics then has an alternate interpretation as the worldsheet of a macroscopic string (see for instance~\rcite{Martinec:1996ad} for a review); gravitational dressing amounts to solving the physical state constraints of the string.  These constraints amount to the projection of the causal structure in field space onto the worldsheet, and physical observables can be written in terms of so-called DDF operators~\rcite{Schoutens:1993hu} (taking the matter sector to consist of free scalar fields~$f^i$)
\begin{align}
\sfA^i(p_+) = \int \!\frac{du}{2\pi} \, e^{ip_+ X^+} \partial_u f
~~,~~~~
\tilde\sfA^i(p_-) = \int \!\frac{dv}{2\pi} \, e^{ip_- X^-} \partial_v f  ~,
\end{align}
where $u,v$ are 2d null coordinates.  These modes were shown to have an exchange algebra
\be
\tilde\sfA^i(p_-)\, \sfA^j(p_+) = e^{-\frac{i}{2\pi} p_+ p_-} \sfA^j(p_+) \, \tilde\sfA^i(p_-) +R_{ij} (p_+,p_-)  ~.
\ee

Initially, it was thought that the non-commutativity of the in- and out-mode operators pointed towards a resolution of the information paradox in terms of complementarity between these sets of observables~\rcite{Kiem:1995iy}.  In hindsight, however, what was observed is simply the 1+1d shock wave S-matrix related to quantum chaos near the horizon~\rcite{Shenker:2013pqa,Maldacena:2015waa,Lam:2018pvp}; information is still lost in these toy models (see for instance~\rcite{Mertens:2019bvy,Moitra:2019xoj} for a recent discussion) because they obey the small corrections theorem and have no compensating non-locality in the field space (\ie\ the target space in the macroscopic string analogy) where the physical dynamics takes place.

Finally, we come to the possibility \ref{code6} that the duality is only an approximation, and that while the CFT is a non-perturbative description of quantum gravity, the bulk theory is not.  This supposition opens up a larger set of possibilities for the encoding of bulk AdS physics in the CFT.  In particular the embedding of Hawking pairs in the Hilbert space of the exact description might not be an isometry,%
\footnote{It should be noted, however, that there is no evidence that AdS/CFT is only approximate rather than an exact duality.  In fact, the impressive matching of bulk computations of supersymmetric black hole partition functions as well as the matching of complicated supergravity solutions to precise states in the dual CFT (see for example~\rcite{Lunin:2001fv,Lin:2004nb,Dabholkar:2012nd,Dabholkar:2014ema,Cabo-Bizet:2018ehj,Benini:2020gjh}) seems to suggest otherwise.}  
and it is suggested that the entangled Hawking state is not exact but rather there is a state-dependent embedding of it and the black hole interior into the exact description, that leads to the island prescription.
For instance, it was pointed out in~\rcite{Akers:2021fut} that one could embed all the product states $\ket{\psi_i}$ of $n$ qubits in a Hilbert space of $m \sim \log n$ qubits in such a way that one captures all their inner products $\bra{\psi_i}\!\psi_j\rangle$ to a high degree of accuracy.  Since one can always write operators in terms of such a product basis
\be
\cO = \sum_{ij} |\psi_i\rangle \cO_{ij} \langle\psi_j|  ~,
\ee
so long as there are not too many off-diagonal terms in the sum (\ie\ the operator acts within a ``code subspace'' of the effective theory of approximately product states), one can well approximate such bulk operators in the exact description, because the evaluation of correlation functions reduces to sums over inner products that are well approximated in the much smaller Hilbert space of the exact theory.  Of course, for states that are not close to product states, and operators that are not close to diagonal in the product state basis,  matrix elements could involve enough terms in the product state basis that the errors in the approximation of the product states accumulate to the point that the bulk description (characterized by these approximate product states) breaks down.

Even so, once again the effective small corrections theorem bypasses these complications, and constrains what one might accomplish by giving up on exact duality between the bulk and boundary descriptions.  All that is needed is that the exact description realizes the product state $\ket{\psi_{\rm pair}}^{\otimes \frac n2}$ with high fidelity, as it must if there is bulk effective field theory and the Hawking process is taking place at the horizon, up to small corrections.  Then the effective small corrections theorem applies and tells us that there are entangled logical qubits involving the radiation quanta $\sfb$ which lead to a monotonically increasing von Neumann entropy of that radiation, absent significant, detectable non-localities in the bulk dynamics.  As in the discussion of~\ref{code1}-\ref{code2} above, these various complexities, approximations and non-localities of the bulk-boundary map are a distraction from the problem at hand.


\section{Discussion} 
\label{sec:discussion}

All the problems reviewed above trace back to the Hawking process and the desire to maintain effective field theory near the horizon.
Every Hawking pair created only adds to the problem we need to solve, because Hawking pairs are created in a unique state and therefore cannot carry information away from the black hole; instead, they generate additional entanglement entropy that must be disentangled by the end of the evaporation process.

The fundamental difference between the Hawking process and radiation from an ordinary body is that in the latter process the state of the radiated quanta depends on the state of the radiating object; in the notation we have been using, the $\sfa$ qubits are being emitted directly, rather than some random $\sfb$ qubits whose state is totally uncorrelated to and unentangled with that of the initial object because they are fully entangled with their partner $\sfc$ qubits.  The (effective) small corrections theorem tells us that the game is over as soon as we postulate the effective field theory Hawking process as the means by which black holes radiate, provided that there is no subsequent non-local dynamical effect that reaches out and modifies the radiation state in some way that depends on the interior state of the black hole.  

There is also the Occam's razor question of why does the system bother with the Hawking process to begin with, if AdS/CFT is capable of the sorts of non-localities needed to patch up the Hawking process; why not simply radiate the information directly using such non-local processes and forego the Hawking mechanism altogether?  Why can't the non-locality support the storage of information at the horizon scale, and thus the ordinary radiation of that stored information from the horizon scale into the bulk?

The problem does not arise if black holes radiate as ordinary bodies do, but this requires some horizon scale microstate structure that can directly radiate $\sfa$ qubits in the way that ordinary bodies do from their surface.  In the context of AdS/CFT, the boundary interaction~\eqref{Hint} then directly transfers $\sfa$ qubits to the bath.  The Page curve then arises as it does for ordinary bodies~-- if the initial state is pure, then the radiation of $\sfa$ qubits first causes the entanglement entropy between the radiation and the remaining hole to rise, until half the $\sfa$ qubits are radiated, and then the entanglement entropy starts to fall because the subsequently radiated qubits are mostly entangled with the ones radiated earlier.

This logic in part motivates the fuzzball proposal.  Rather than invoking non-local processes that act on arbitrarily distant degrees of freedom, in the fuzzball scenario one only requires effects that act at the horizon scale.  More precisely, rather than non-locality, the fuzzball scenario proposes that the quantum wavefunction of the black hole constituents is quantum coherent over the horizon scale.  One gives up the Hawking process at the horizon, as it is the source of all the subsequent difficulties we have seen above.

One might then ask, absent the Hawking process, how does one manage to recover all the usual thermodynamic properties of black holes?  Naively it seems that is the structure of the vacuum geometry and its smoothness at the horizon that lead to those properties.  This includes the calculation of the black hole temperature that results from the Hawking process.  Don't we lose all this structure if we give up vacuum physics at the horizon?  

The answer to this question lies in the very emergence of gravity as the effective theory {\it outside the horizon}.  It is the effective gravity theory that determines the equation of state \eg\ via the smoothness of the Euclidean continuation of the exterior geometry~\rcite{Gibbons:1976ue}; or alternatively via the boost symmetry of the exterior static geometry~\rcite{Wald:1993nt} in Lorentz signature.  Ultimately the thermodynamic properties arise from the emergent diffeomorphism symmetry of the effective theory.   The entropy is a Noether charge related to the near-horizon boost symmetry; the temperature is given by the surface gravity and thus also determined by the symmetry.

Whatever the corrections to the emergent low-energy effective theory and its diffeomorphism symmetry, it is likely that those effects are confined to a small ``stretched horizon'' that extends slightly outside the horizon of the black hole solution of the effective theory.  One can then compare the exact theory which includes those corrections to the solution of the effective theory (\eg\ supergravity), and they will match closely outside the stretched horizon.  For instance, in the Euclidean solution we would be cutting a small disc out of the $r$-$t$ plane near the Euclidean continuation of the black hole horizon, and the exact and supergravity solutions match well in the region outside that disk.  Filling in that disk with a smooth geometry solving the supergravity field equations, and thereby extrapolating the effective supergravity geometry all the way to its horizon, makes a mistake relative to the exact theory that is small when the black hole is sufficiently large and semi-classical.

An illustrative example of this behavior is the black fivebrane, whose Euclidean geometry takes the form of the gauged WZW model $\frac\sltwo\uone$ in the $r$-$t$ plane
\be
ds^2 = k\big(dr^2 + \tanh^2\!r\,dt^2\big)
\ee
(where $k$ is the level of the underlying $\sltwo$ current algebra of worldsheet string theory).
This geometry is inextricably linked to a stringy condensate concentrated within a string length of $r=0$, whose properties are precisely known due to a property of the coset model known as FZZ duality (see for instance~\rcite{Giveon:2016dxe} for an overview), yet this feature does not affect the thermodynamics at leading order~\rcite{Kutasov:2000jp,Giveon:2005jv} in the large $k$ limit.

In gauge/gravity duality, the onset of the black hole regime is associated to a Hawking/Page phase transition on the gravity side, dual to a deconfinement transition on the gauge theory side.  The obvious candidate for horizon scale ``fuzz'' is the deconfined phase, which would shut off at the horizon scale but characterizes the black hole interior.  The black hole would then be a compact object composed of this exotic phase of matter, radiating like an ordinary body.  Its entropy, temperature, and other thermodynamic properties are guaranteed to match those predicted by general relativity because the exterior geometry is unchanged (up to small corrections), and the symmetries of that effective geometry determine the thermodynamics regardless of what exotic phase governs the black hole interior.

If one is giving up vacuum physics at the horizon scale, a natural followup question concerns the experience of the infalling observer who crosses the horizon into the fuzzball.  This is a question about the response function of ``fuzz'' to probes of high transverse momentum.  We don't know enough about the brane dynamics that constitutes ``fuzz'' in the relevant dense and strongly coupled regime in order to answer this question.  The ``firewall'' scenario~\rcite{Almheiri:2012rt} argues for a stiff response function; the idea of ``fuzzball complementarity''~\rcite{Mathur:2010kx,Mathur:2012jk,Avery:2012tf} suggests the possibility of a soft response function at high transverse momenta.  The latter idea accords with what we know of string theory at weak coupling and low density, where large transverse momentum exchanges over short periods of time are avoided~\rcite{Gross:1987ar}.  This soft response scenario leads to a sort of ``bag model'' picture of the quantum black hole, in which a confined exterior (whose excitations are gauge singlet supergravitons) makes an abrupt transition to a deconfined interior.  A probe crossing the phase boundary might not experience strong kicks but instead fragment over some penetration depth over which it scrambles and eventually thermalizes (like a QCD meson entering a quark-gluon plasma and forming a jet).  The debate about firewalls versus fuzzball complementerity is  essentially the question of what is the penetration depth relative to the horizon scale.  However, we should stress that these are simply informed speculations as to the nature of ``fuzz''; one hopes that someday such speculations will be supplanted by reliable calculations.




\section*{Acknowledgements}

I thank 
G. Penington,
M. Rangamani,
and especially
S. Mathur
for discussions.
This work is supported in part by DOE grant DE-SC0009924.


\appendix


\vskip 3cm

\bibliographystyle{JHEP}      

\bibliography{fivebranes}

\providecommand{\href}[2]{#2}\begingroup\raggedright\begin{thebibliography}{10}

\bibitem{Strominger:1996sh}
A.~Strominger and C.~Vafa, \emph{{Microscopic origin of the Bekenstein-Hawking
  entropy}}, \href{http://dx.doi.org/10.1016/0370-2693(96)00345-0}{\emph{Phys.
  Lett. B} {\bfseries 379} (1996) 99--104},
  [\href{https://arxiv.org/abs/hep-th/9601029}{{\ttfamily hep-th/9601029}}].

\bibitem{Susskind:1993if}
L.~Susskind, L.~Thorlacius and J.~Uglum, \emph{{The Stretched horizon and black
  hole complementarity}},
  \href{http://dx.doi.org/10.1103/PhysRevD.48.3743}{\emph{Phys. Rev. D}
  {\bfseries 48} (1993) 3743--3761},
  [\href{https://arxiv.org/abs/hep-th/9306069}{{\ttfamily hep-th/9306069}}].

\bibitem{Banks:1996vh}
T.~Banks, W.~Fischler, S.~H. Shenker and L.~Susskind, \emph{{M theory as a
  matrix model: A Conjecture}},
  \href{http://dx.doi.org/10.1103/PhysRevD.55.5112}{\emph{Phys. Rev. D}
  {\bfseries 55} (1997) 5112--5128},
  [\href{https://arxiv.org/abs/hep-th/9610043}{{\ttfamily hep-th/9610043}}].

\bibitem{Mathur:2009hf}
S.~D. Mathur, \emph{{The Information paradox: A Pedagogical introduction}},
  \href{http://dx.doi.org/10.1088/0264-9381/26/22/224001}{\emph{Class. Quant.
  Grav.} {\bfseries 26} (2009) 224001},
  [\href{https://arxiv.org/abs/0909.1038}{{\ttfamily 0909.1038}}].

\bibitem{Braunstein:2009my}
S.~L. Braunstein, S.~Pirandola and K.~\.Zyczkowski, \emph{{Better Late than
  Never: Information Retrieval from Black Holes}},
  \href{http://dx.doi.org/10.1103/PhysRevLett.110.101301}{\emph{Phys. Rev.
  Lett.} {\bfseries 110} (2013) 101301},
  [\href{https://arxiv.org/abs/0907.1190}{{\ttfamily 0907.1190}}].

\bibitem{Almheiri:2012rt}
A.~Almheiri, D.~Marolf, J.~Polchinski and J.~Sully, \emph{{Black Holes:
  Complementarity or Firewalls?}},
  \href{http://dx.doi.org/10.1007/JHEP02(2013)062}{\emph{JHEP} {\bfseries 02}
  (2013) 062}, [\href{https://arxiv.org/abs/1207.3123}{{\ttfamily 1207.3123}}].

\bibitem{Maldacena:2001kr}
J.~M. Maldacena, \emph{{Eternal black holes in anti-de Sitter}},
  \href{http://dx.doi.org/10.1088/1126-6708/2003/04/021}{\emph{JHEP} {\bfseries
  04} (2003) 021}, [\href{https://arxiv.org/abs/hep-th/0106112}{{\ttfamily
  hep-th/0106112}}].

\bibitem{Ryu:2006bv}
S.~Ryu and T.~Takayanagi, \emph{{Holographic derivation of entanglement entropy
  from AdS/CFT}},
  \href{http://dx.doi.org/10.1103/PhysRevLett.96.181602}{\emph{Phys. Rev.
  Lett.} {\bfseries 96} (2006) 181602},
  [\href{https://arxiv.org/abs/hep-th/0603001}{{\ttfamily hep-th/0603001}}].

\bibitem{VanRaamsdonk:2010pw}
M.~Van~Raamsdonk, \emph{{Building up spacetime with quantum entanglement}},
  \href{http://dx.doi.org/10.1142/S0218271810018529}{\emph{Gen. Rel. Grav.}
  {\bfseries 42} (2010) 2323--2329},
  [\href{https://arxiv.org/abs/1005.3035}{{\ttfamily 1005.3035}}].

\bibitem{Engelhardt:2014gca}
N.~Engelhardt and A.~C. Wall, \emph{{Quantum Extremal Surfaces: Holographic
  Entanglement Entropy beyond the Classical Regime}},
  \href{http://dx.doi.org/10.1007/JHEP01(2015)073}{\emph{JHEP} {\bfseries 01}
  (2015) 073}, [\href{https://arxiv.org/abs/1408.3203}{{\ttfamily 1408.3203}}].

\bibitem{Penington:2019npb}
G.~Penington, \emph{{Entanglement Wedge Reconstruction and the Information
  Paradox}}, \href{http://dx.doi.org/10.1007/JHEP09(2020)002}{\emph{JHEP}
  {\bfseries 09} (2020) 002},
  [\href{https://arxiv.org/abs/1905.08255}{{\ttfamily 1905.08255}}].

\bibitem{Almheiri:2019psf}
A.~Almheiri, N.~Engelhardt, D.~Marolf and H.~Maxfield, \emph{{The entropy of
  bulk quantum fields and the entanglement wedge of an evaporating black
  hole}}, \href{http://dx.doi.org/10.1007/JHEP12(2019)063}{\emph{JHEP}
  {\bfseries 12} (2019) 063},
  [\href{https://arxiv.org/abs/1905.08762}{{\ttfamily 1905.08762}}].

\bibitem{Almheiri:2013hfa}
A.~Almheiri, D.~Marolf, J.~Polchinski, D.~Stanford and J.~Sully, \emph{{An
  Apologia for Firewalls}},
  \href{http://dx.doi.org/10.1007/JHEP09(2013)018}{\emph{JHEP} {\bfseries 09}
  (2013) 018}, [\href{https://arxiv.org/abs/1304.6483}{{\ttfamily 1304.6483}}].

\bibitem{Guo:2021blh}
B.~Guo, M.~R.~R. Hughes, S.~D. Mathur and M.~Mehta, \emph{{Contrasting the
  fuzzball and wormhole paradigms for black holes}},
  \href{https://arxiv.org/abs/2111.05295}{{\ttfamily 2111.05295}}.

\bibitem{Wald:2019ygd}
R.~M. Wald, \emph{{Particle and energy cost of entanglement of Hawking
  radiation with the final vacuum state}},
  \href{http://dx.doi.org/10.1103/PhysRevD.100.065019}{\emph{Phys. Rev. D}
  {\bfseries 100} (2019) 065019},
  [\href{https://arxiv.org/abs/1908.06363}{{\ttfamily 1908.06363}}].

\bibitem{Callan:1996dv}
C.~G. Callan and J.~M. Maldacena, \emph{{D-brane approach to black hole quantum
  mechanics}},
  \href{http://dx.doi.org/10.1016/0550-3213(96)00225-8}{\emph{Nucl. Phys. B}
  {\bfseries 472} (1996) 591--610},
  [\href{https://arxiv.org/abs/hep-th/9602043}{{\ttfamily hep-th/9602043}}].

\bibitem{Das:1996wn}
S.~R. Das and S.~D. Mathur, \emph{{Comparing decay rates for black holes and
  D-branes}}, \href{http://dx.doi.org/10.1016/0550-3213(96)00453-1}{\emph{Nucl.
  Phys. B} {\bfseries 478} (1996) 561--576},
  [\href{https://arxiv.org/abs/hep-th/9606185}{{\ttfamily hep-th/9606185}}].

\bibitem{Maldacena:1996ix}
J.~M. Maldacena and A.~Strominger, \emph{{Black hole grey body factors and
  d-brane spectroscopy}},
  \href{http://dx.doi.org/10.1103/PhysRevD.55.861}{\emph{Phys. Rev. D}
  {\bfseries 55} (1997) 861--870},
  [\href{https://arxiv.org/abs/hep-th/9609026}{{\ttfamily hep-th/9609026}}].

\bibitem{Chowdhury:2007jx}
B.~D. Chowdhury and S.~D. Mathur, \emph{{Radiation from the non-extremal
  fuzzball}},
  \href{http://dx.doi.org/10.1088/0264-9381/25/13/135005}{\emph{Class. Quant.
  Grav.} {\bfseries 25} (2008) 135005},
  [\href{https://arxiv.org/abs/0711.4817}{{\ttfamily 0711.4817}}].

\bibitem{Chowdhury:2008uj}
B.~D. Chowdhury and S.~D. Mathur, \emph{{Non-extremal fuzzballs and ergoregion
  emission}},
  \href{http://dx.doi.org/10.1088/0264-9381/26/3/035006}{\emph{Class. Quant.
  Grav.} {\bfseries 26} (2009) 035006},
  [\href{https://arxiv.org/abs/0810.2951}{{\ttfamily 0810.2951}}].

\bibitem{Avery:2009tu}
S.~G. Avery, B.~D. Chowdhury and S.~D. Mathur, \emph{{Emission from the D1D5
  CFT}}, \href{http://dx.doi.org/10.1088/1126-6708/2009/10/065}{\emph{JHEP}
  {\bfseries 10} (2009) 065},
  [\href{https://arxiv.org/abs/0906.2015}{{\ttfamily 0906.2015}}].

\bibitem{Giddings:2009ae}
S.~B. Giddings, \emph{{Nonlocality versus complementarity: A Conservative
  approach to the information problem}},
  \href{http://dx.doi.org/10.1088/0264-9381/28/2/025002}{\emph{Class. Quant.
  Grav.} {\bfseries 28} (2011) 025002},
  [\href{https://arxiv.org/abs/0911.3395}{{\ttfamily 0911.3395}}].

\bibitem{Giddings:2011ks}
S.~B. Giddings, \emph{{Models for unitary black hole disintegration}},
  \href{http://dx.doi.org/10.1103/PhysRevD.85.044038}{\emph{Phys. Rev. D}
  {\bfseries 85} (2012) 044038},
  [\href{https://arxiv.org/abs/1108.2015}{{\ttfamily 1108.2015}}].

\bibitem{Giddings:2012gc}
S.~B. Giddings, \emph{{Nonviolent nonlocality}},
  \href{http://dx.doi.org/10.1103/PhysRevD.88.064023}{\emph{Phys. Rev. D}
  {\bfseries 88} (2013) 064023},
  [\href{https://arxiv.org/abs/1211.7070}{{\ttfamily 1211.7070}}].

\bibitem{Giddings:2013kcj}
S.~B. Giddings, \emph{{Nonviolent information transfer from black holes: A
  field theory parametrization}},
  \href{http://dx.doi.org/10.1103/PhysRevD.88.024018}{\emph{Phys. Rev. D}
  {\bfseries 88} (2013) 024018},
  [\href{https://arxiv.org/abs/1302.2613}{{\ttfamily 1302.2613}}].

\bibitem{Giddings:2013noa}
S.~B. Giddings and Y.~Shi, \emph{{Effective field theory models for nonviolent
  information transfer from black holes}},
  \href{http://dx.doi.org/10.1103/PhysRevD.89.124032}{\emph{Phys. Rev. D}
  {\bfseries 89} (2014) 124032},
  [\href{https://arxiv.org/abs/1310.5700}{{\ttfamily 1310.5700}}].

\bibitem{Giddings:2014nla}
S.~B. Giddings, \emph{{Modulated Hawking radiation and a nonviolent channel for
  information release}},
  \href{http://dx.doi.org/10.1016/j.physletb.2014.08.070}{\emph{Phys. Lett. B}
  {\bfseries 738} (2014) 92--96},
  [\href{https://arxiv.org/abs/1401.5804}{{\ttfamily 1401.5804}}].

\bibitem{Giddings:2017mym}
S.~B. Giddings, \emph{{Nonviolent unitarization: basic postulates to soft
  quantum structure of black holes}},
  \href{http://dx.doi.org/10.1007/JHEP12(2017)047}{\emph{JHEP} {\bfseries 12}
  (2017) 047}, [\href{https://arxiv.org/abs/1701.08765}{{\ttfamily
  1701.08765}}].

\bibitem{Giddings:2012dh}
S.~B. Giddings and Y.~Shi, \emph{{Quantum information transfer and models for
  black hole mechanics}},
  \href{http://dx.doi.org/10.1103/PhysRevD.87.064031}{\emph{Phys. Rev. D}
  {\bfseries 87} (2013) 064031},
  [\href{https://arxiv.org/abs/1205.4732}{{\ttfamily 1205.4732}}].

\bibitem{Giddings:2013vda}
S.~B. Giddings, \emph{{Statistical physics of black holes as quantum-mechanical
  systems}}, \href{http://dx.doi.org/10.1103/PhysRevD.88.104013}{\emph{Phys.
  Rev. D} {\bfseries 88} (2013) 104013},
  [\href{https://arxiv.org/abs/1308.3488}{{\ttfamily 1308.3488}}].

\bibitem{Cvetic:1997uw}
M.~Cvetic and F.~Larsen, \emph{{General rotating black holes in string theory:
  Grey body factors and event horizons}},
  \href{http://dx.doi.org/10.1103/PhysRevD.56.4994}{\emph{Phys. Rev. D}
  {\bfseries 56} (1997) 4994--5007},
  [\href{https://arxiv.org/abs/hep-th/9705192}{{\ttfamily hep-th/9705192}}].

\bibitem{Klebanov:1997cx}
I.~R. Klebanov and S.~D. Mathur, \emph{{Black hole grey body factors and
  absorption of scalars by effective strings}},
  \href{http://dx.doi.org/10.1016/S0550-3213(97)00287-3}{\emph{Nucl. Phys. B}
  {\bfseries 500} (1997) 115--132},
  [\href{https://arxiv.org/abs/hep-th/9701187}{{\ttfamily hep-th/9701187}}].

\bibitem{Hawking:2016msc}
S.~W. Hawking, M.~J. Perry and A.~Strominger, \emph{{Soft Hair on Black
  Holes}}, \href{http://dx.doi.org/10.1103/PhysRevLett.116.231301}{\emph{Phys.
  Rev. Lett.} {\bfseries 116} (2016) 231301},
  [\href{https://arxiv.org/abs/1601.00921}{{\ttfamily 1601.00921}}].

\bibitem{Papadodimas:2013wnh}
K.~Papadodimas and S.~Raju, \emph{{Black Hole Interior in the Holographic
  Correspondence and the Information Paradox}},
  \href{http://dx.doi.org/10.1103/PhysRevLett.112.051301}{\emph{Phys. Rev.
  Lett.} {\bfseries 112} (2014) 051301},
  [\href{https://arxiv.org/abs/1310.6334}{{\ttfamily 1310.6334}}].

\bibitem{Papadodimas:2013jku}
K.~Papadodimas and S.~Raju, \emph{{State-Dependent Bulk-Boundary Maps and Black
  Hole Complementarity}},
  \href{http://dx.doi.org/10.1103/PhysRevD.89.086010}{\emph{Phys. Rev. D}
  {\bfseries 89} (2014) 086010},
  [\href{https://arxiv.org/abs/1310.6335}{{\ttfamily 1310.6335}}].

\bibitem{Susskind:2013tg}
L.~Susskind, \emph{{Black Hole Complementarity and the Harlow-Hayden
  Conjecture}},  \href{https://arxiv.org/abs/1301.4505}{{\ttfamily 1301.4505}}.

\bibitem{Maldacena:2013xja}
J.~Maldacena and L.~Susskind, \emph{{Cool horizons for entangled black holes}},
  \href{http://dx.doi.org/10.1002/prop.201300020}{\emph{Fortsch. Phys.}
  {\bfseries 61} (2013) 781--811},
  [\href{https://arxiv.org/abs/1306.0533}{{\ttfamily 1306.0533}}].

\bibitem{Verlinde:2013qya}
E.~Verlinde and H.~Verlinde, \emph{{Behind the Horizon in AdS/CFT}},
  \href{https://arxiv.org/abs/1311.1137}{{\ttfamily 1311.1137}}.

\bibitem{Harlow:2014yoa}
D.~Harlow, \emph{{Aspects of the Papadodimas-Raju Proposal for the Black Hole
  Interior}}, \href{http://dx.doi.org/10.1007/JHEP11(2014)055}{\emph{JHEP}
  {\bfseries 11} (2014) 055},
  [\href{https://arxiv.org/abs/1405.1995}{{\ttfamily 1405.1995}}].

\bibitem{Harlow:2014yka}
D.~Harlow, \emph{{Jerusalem Lectures on Black Holes and Quantum Information}},
  \href{http://dx.doi.org/10.1103/RevModPhys.88.015002}{\emph{Rev. Mod. Phys.}
  {\bfseries 88} (2016) 015002},
  [\href{https://arxiv.org/abs/1409.1231}{{\ttfamily 1409.1231}}].

\bibitem{Marolf:2015dia}
D.~Marolf and J.~Polchinski, \emph{{Violations of the Born rule in cool
  state-dependent horizons}},
  \href{http://dx.doi.org/10.1007/JHEP01(2016)008}{\emph{JHEP} {\bfseries 01}
  (2016) 008}, [\href{https://arxiv.org/abs/1506.01337}{{\ttfamily
  1506.01337}}].

\bibitem{Papadodimas:2015jra}
K.~Papadodimas and S.~Raju, \emph{{Remarks on the necessity and implications of
  state-dependence in the black hole interior}},
  \href{http://dx.doi.org/10.1103/PhysRevD.93.084049}{\emph{Phys. Rev. D}
  {\bfseries 93} (2016) 084049},
  [\href{https://arxiv.org/abs/1503.08825}{{\ttfamily 1503.08825}}].

\bibitem{Papadodimas:2017qit}
K.~Papadodimas, \emph{{A class of non-equilibrium states and the black hole
  interior}},  \href{https://arxiv.org/abs/1708.06328}{{\ttfamily 1708.06328}}.

\bibitem{Kim:2020cds}
I.~Kim, E.~Tang and J.~Preskill, \emph{{The ghost in the radiation: Robust
  encodings of the black hole interior}},
  \href{http://dx.doi.org/10.1007/JHEP06(2020)031}{\emph{JHEP} {\bfseries 06}
  (2020) 031}, [\href{https://arxiv.org/abs/2003.05451}{{\ttfamily
  2003.05451}}].

\bibitem{Gao:2016bin}
P.~Gao, D.~L. Jafferis and A.~C. Wall, \emph{{Traversable Wormholes via a
  Double Trace Deformation}},
  \href{http://dx.doi.org/10.1007/JHEP12(2017)151}{\emph{JHEP} {\bfseries 12}
  (2017) 151}, [\href{https://arxiv.org/abs/1608.05687}{{\ttfamily
  1608.05687}}].

\bibitem{Maldacena:2018lmt}
J.~Maldacena and X.-L. Qi, \emph{{Eternal traversable wormhole}},
  \href{https://arxiv.org/abs/1804.00491}{{\ttfamily 1804.00491}}.

\bibitem{Raju:2020smc}
S.~Raju, \emph{{Lessons from the information paradox}},
  \href{http://dx.doi.org/10.1016/j.physrep.2021.10.001}{\emph{Phys. Rept.}
  {\bfseries 943} (2022) 2187},
  [\href{https://arxiv.org/abs/2012.05770}{{\ttfamily 2012.05770}}].

\bibitem{Raju:2021lwh}
S.~Raju, \emph{{Failure of the split property in gravity and the information
  paradox}},  \href{https://arxiv.org/abs/2110.05470}{{\ttfamily 2110.05470}}.

\bibitem{Donnelly:2018nbv}
W.~Donnelly and S.~B. Giddings, \emph{{Gravitational splitting at first order:
  Quantum information localization in gravity}},
  \href{http://dx.doi.org/10.1103/PhysRevD.98.086006}{\emph{Phys. Rev. D}
  {\bfseries 98} (2018) 086006},
  [\href{https://arxiv.org/abs/1805.11095}{{\ttfamily 1805.11095}}].

\bibitem{Giddings:2021khn}
S.~B. Giddings, \emph{{On the questions of asymptotic recoverability of
  information and subsystems in quantum gravity}},
  \href{https://arxiv.org/abs/2112.03207}{{\ttfamily 2112.03207}}.

\bibitem{Balasubramanian:2005qu}
V.~Balasubramanian, P.~Kraus and M.~Shigemori, \emph{{Massless black holes and
  black rings as effective geometries of the D1-D5 system}},
  \href{http://dx.doi.org/10.1088/0264-9381/22/22/010}{\emph{Class. Quant.
  Grav.} {\bfseries 22} (2005) 4803--4838},
  [\href{https://arxiv.org/abs/hep-th/0508110}{{\ttfamily hep-th/0508110}}].

\bibitem{Balasubramanian:2018yjq}
V.~Balasubramanian, D.~Berenstein, A.~Lewkowycz, A.~Miller, O.~Parrikar and
  C.~Rabideau, \emph{{Emergent classical spacetime from microstates of an
  incipient black hole}},
  \href{http://dx.doi.org/10.1007/JHEP01(2019)197}{\emph{JHEP} {\bfseries 01}
  (2019) 197}, [\href{https://arxiv.org/abs/1810.13440}{{\ttfamily
  1810.13440}}].

\bibitem{Martinec:1996ad}
E.~J. Martinec, \emph{{Two-dimensional models of black hole evaporation}},
  \href{http://dx.doi.org/10.1088/0264-9381/13/1/003}{\emph{Class. Quant.
  Grav.} {\bfseries 13} (1996) 1--25}.

\bibitem{Schoutens:1993hu}
K.~Schoutens, H.~L. Verlinde and E.~P. Verlinde, \emph{{Quantum black hole
  evaporation}}, \href{http://dx.doi.org/10.1103/PhysRevD.48.2670}{\emph{Phys.
  Rev. D} {\bfseries 48} (1993) 2670--2685},
  [\href{https://arxiv.org/abs/hep-th/9304128}{{\ttfamily hep-th/9304128}}].

\bibitem{Kiem:1995iy}
Y.~Kiem, H.~L. Verlinde and E.~P. Verlinde, \emph{{Black hole horizons and
  complementarity}},
  \href{http://dx.doi.org/10.1103/PhysRevD.52.7053}{\emph{Phys. Rev. D}
  {\bfseries 52} (1995) 7053--7065},
  [\href{https://arxiv.org/abs/hep-th/9502074}{{\ttfamily hep-th/9502074}}].

\bibitem{Shenker:2013pqa}
S.~H. Shenker and D.~Stanford, \emph{{Black holes and the butterfly effect}},
  \href{http://dx.doi.org/10.1007/JHEP03(2014)067}{\emph{JHEP} {\bfseries 03}
  (2014) 067}, [\href{https://arxiv.org/abs/1306.0622}{{\ttfamily 1306.0622}}].

\bibitem{Maldacena:2015waa}
J.~Maldacena, S.~H. Shenker and D.~Stanford, \emph{{A bound on chaos}},
  \href{http://dx.doi.org/10.1007/JHEP08(2016)106}{\emph{JHEP} {\bfseries 08}
  (2016) 106}, [\href{https://arxiv.org/abs/1503.01409}{{\ttfamily
  1503.01409}}].

\bibitem{Lam:2018pvp}
H.~T. Lam, T.~G. Mertens, G.~J. Turiaci and H.~Verlinde, \emph{{Shockwave
  S-matrix from Schwarzian Quantum Mechanics}},
  \href{http://dx.doi.org/10.1007/JHEP11(2018)182}{\emph{JHEP} {\bfseries 11}
  (2018) 182}, [\href{https://arxiv.org/abs/1804.09834}{{\ttfamily
  1804.09834}}].

\bibitem{Mertens:2019bvy}
T.~G. Mertens, \emph{{Towards Black Hole Evaporation in Jackiw-Teitelboim
  Gravity}}, \href{http://dx.doi.org/10.1007/JHEP07(2019)097}{\emph{JHEP}
  {\bfseries 07} (2019) 097},
  [\href{https://arxiv.org/abs/1903.10485}{{\ttfamily 1903.10485}}].

\bibitem{Moitra:2019xoj}
U.~Moitra, S.~K. Sake, S.~P. Trivedi and V.~Vishal, \emph{{Jackiw-Teitelboim
  Model Coupled to Conformal Matter in the Semi-Classical Limit}},
  \href{http://dx.doi.org/10.1007/JHEP04(2020)199}{\emph{JHEP} {\bfseries 04}
  (2020) 199}, [\href{https://arxiv.org/abs/1908.08523}{{\ttfamily
  1908.08523}}].

\bibitem{Lunin:2001fv}
O.~Lunin and S.~D. Mathur, \emph{{Metric of the multiply wound rotating
  string}}, \href{http://dx.doi.org/10.1016/S0550-3213(01)00321-2}{\emph{Nucl.
  Phys.} {\bfseries B610} (2001) 49--76},
  [\href{https://arxiv.org/abs/hep-th/0105136}{{\ttfamily hep-th/0105136}}].

\bibitem{Lin:2004nb}
H.~Lin, O.~Lunin and J.~M. Maldacena, \emph{{Bubbling AdS space and 1/2 BPS
  geometries}},
  \href{http://dx.doi.org/10.1088/1126-6708/2004/10/025}{\emph{JHEP} {\bfseries
  10} (2004) 025}, [\href{https://arxiv.org/abs/hep-th/0409174}{{\ttfamily
  hep-th/0409174}}].

\bibitem{Dabholkar:2012nd}
A.~Dabholkar, S.~Murthy and D.~Zagier, \emph{{Quantum Black Holes, Wall
  Crossing, and Mock Modular Forms}},
  \href{https://arxiv.org/abs/1208.4074}{{\ttfamily 1208.4074}}.

\bibitem{Dabholkar:2014ema}
A.~Dabholkar, J.~Gomes and S.~Murthy, \emph{{Nonperturbative black hole entropy
  and Kloosterman sums}},
  \href{http://dx.doi.org/10.1007/JHEP03(2015)074}{\emph{JHEP} {\bfseries 03}
  (2015) 074}, [\href{https://arxiv.org/abs/1404.0033}{{\ttfamily 1404.0033}}].

\bibitem{Cabo-Bizet:2018ehj}
A.~Cabo-Bizet, D.~Cassani, D.~Martelli and S.~Murthy, \emph{{Microscopic origin
  of the Bekenstein-Hawking entropy of supersymmetric AdS$_{5}$ black holes}},
  \href{http://dx.doi.org/10.1007/JHEP10(2019)062}{\emph{JHEP} {\bfseries 10}
  (2019) 062}, [\href{https://arxiv.org/abs/1810.11442}{{\ttfamily
  1810.11442}}].

\bibitem{Benini:2020gjh}
F.~Benini, E.~Colombo, S.~Soltani, A.~Zaffaroni and Z.~Zhang,
  \emph{{Superconformal indices at large $N$ and the entropy of AdS$_5$
  $\times$ SE$_5$ black holes}},
  \href{http://dx.doi.org/10.1088/1361-6382/abb39b}{\emph{Class. Quant. Grav.}
  {\bfseries 37} (2020) 215021},
  [\href{https://arxiv.org/abs/2005.12308}{{\ttfamily 2005.12308}}].

\bibitem{Akers:2021fut}
C.~Akers and G.~Penington, \emph{{Quantum minimal surfaces from quantum error
  correction}},  \href{https://arxiv.org/abs/2109.14618}{{\ttfamily
  2109.14618}}.

\bibitem{Gibbons:1976ue}
G.~W. Gibbons and S.~W. Hawking, \emph{{Action Integrals and Partition
  Functions in Quantum Gravity}},
  \href{http://dx.doi.org/10.1103/PhysRevD.15.2752}{\emph{Phys. Rev. D}
  {\bfseries 15} (1977) 2752--2756}.

\bibitem{Wald:1993nt}
R.~M. Wald, \emph{{Black hole entropy is the Noether charge}},
  \href{http://dx.doi.org/10.1103/PhysRevD.48.R3427}{\emph{Phys. Rev. D}
  {\bfseries 48} (1993) R3427--R3431},
  [\href{https://arxiv.org/abs/gr-qc/9307038}{{\ttfamily gr-qc/9307038}}].

\bibitem{Giveon:2016dxe}
A.~Giveon, N.~Itzhaki and D.~Kutasov, \emph{{Stringy Horizons II}},
  \href{http://dx.doi.org/10.1007/JHEP10(2016)157}{\emph{JHEP} {\bfseries 10}
  (2016) 157}, [\href{https://arxiv.org/abs/1603.05822}{{\ttfamily
  1603.05822}}].

\bibitem{Kutasov:2000jp}
D.~Kutasov and D.~A. Sahakyan, \emph{{Comments on the thermodynamics of little
  string theory}},
  \href{http://dx.doi.org/10.1088/1126-6708/2001/02/021}{\emph{JHEP} {\bfseries
  02} (2001) 021}, [\href{https://arxiv.org/abs/hep-th/0012258}{{\ttfamily
  hep-th/0012258}}].

\bibitem{Giveon:2005jv}
A.~Giveon and D.~Kutasov, \emph{{The Charged black hole/string transition}},
  \href{http://dx.doi.org/10.1088/1126-6708/2006/01/120}{\emph{JHEP} {\bfseries
  01} (2006) 120}, [\href{https://arxiv.org/abs/hep-th/0510211}{{\ttfamily
  hep-th/0510211}}].

\bibitem{Mathur:2010kx}
S.~D. Mathur, \emph{{The Information paradox and the infall problem}},
  \href{http://dx.doi.org/10.1088/0264-9381/28/12/125010}{\emph{Class. Quant.
  Grav.} {\bfseries 28} (2011) 125010},
  [\href{https://arxiv.org/abs/1012.2101}{{\ttfamily 1012.2101}}].

\bibitem{Mathur:2012jk}
S.~D. Mathur and D.~Turton, \emph{{Comments on black holes I: The possibility
  of complementarity}},
  \href{http://dx.doi.org/10.1007/JHEP01(2014)034}{\emph{JHEP} {\bfseries 01}
  (2014) 034}, [\href{https://arxiv.org/abs/1208.2005}{{\ttfamily 1208.2005}}].

\bibitem{Avery:2012tf}
S.~G. Avery, B.~D. Chowdhury and A.~Puhm, \emph{{Unitarity and fuzzball
  complementarity: 'Alice fuzzes but may not even know it!'}},
  \href{http://dx.doi.org/10.1007/JHEP09(2013)012}{\emph{JHEP} {\bfseries 09}
  (2013) 012}, [\href{https://arxiv.org/abs/1210.6996}{{\ttfamily 1210.6996}}].

\bibitem{Gross:1987ar}
D.~J. Gross and P.~F. Mende, \emph{{String Theory Beyond the Planck Scale}},
  \href{http://dx.doi.org/10.1016/0550-3213(88)90390-2}{\emph{Nucl. Phys. B}
  {\bfseries 303} (1988) 407--454}.

\end{thebibliography}\endgroup


\end{document}